\documentclass[12pt,a4paper,final]{iopart}

\usepackage{iopams}
\usepackage{graphicx}
\usepackage[breaklinks=true,colorlinks=true,linkcolor=blue,urlcolor=blue,citecolor=blue]{hyperref}

\expandafter\let\csname equation*\endcsname\relax
\expandafter\let\csname endequation*\endcsname\relax
\usepackage{amsmath,amsfonts}
\usepackage{tikz}
\usepackage{color}
\usetikzlibrary{arrows, automata}
\usepackage[normalem]{ulem}

\newcommand\addi[1]{{\color{black} #1}}
\newcommand\addii[1]{{\color{black} #1}}
\newcommand\deletei{\bgroup\markoverwith{\textcolor{red}{\rule[0.5ex]{2pt}{2pt}}}\ULon}
\newcommand\deleteii{\bgroup\markoverwith{\textcolor{blue}{\rule[0.5ex]{2pt}{2pt}}}\ULon}

\begin{document}

\title[calculation of photoionization in streamer discharges using FMM]{Accurate and efficient calculation of photoionization in streamer discharges using fast multipole method}

\author{Bo Lin}
\address{Department of Mathematics, National University of Singapore, 10 Lower Kent Ridge Road, Singapore 119076}
\ead{linbo@u.nus.edu}

\author{Chijie Zhuang}
\address{State Key Lab of Power Systems and Department of Electrical Engineering, Tsinghua University, Beijing 100084, China}
\ead{chijie@tsinghua.edu.cn}

\author{Zhenning Cai}
\address{Department of Mathematics, National University of Singapore, 10 Lower Kent Ridge Road, Singapore 119076}
\ead{matcz@nus.edu.sg}

\author{Rong Zeng}
\address{State Key Lab of Power Systems and Department of Electrical Engineering, Tsinghua University, Beijing 100084, China}
\ead{zengrong@tsinghua.edu.cn}

\author[cor1]{Weizhu Bao}
\address{Department of Mathematics, National University of Singapore, 10 Lower Kent Ridge Road, Singapore 119076}
\eads{\mailto{matbaowz@nus.edu.sg}}

\begin{abstract}
This paper focuses on the three-dimensional simulation of the photoionization in streamer discharges, and provides a general framework to efficiently and accurately calculate the photoionization model using the integral form. The simulation is based on the kernel-independent fast multipole method. The accuracy of this method is studied quantitatively for different domains and various pressures in comparison with other existing models based on partial differential equations (PDEs). The comparison indicates the numerical error of the fast multipole method is much smaller than those of other PDE-based methods, with the reference solution given by direct numerical integration. Such accuracy can be achieved with affordable computational cost, and its performance in both efficiency and accuracy is quite stable for different domains and pressures. Meanwhile, the simulation accelerated by the fast multipole method exhibits good scalability using up to 1280 cores, which shows its capability of three-dimensional simulations using parallel (distributed) computing. The difference of the proposed method and other efficient approximations are also studied in a three-dimensional dynamic problem where two streamers interact.

\end{abstract}

\vspace{2pc}

%\submitto{\JPD}
%\maketitle

\section{Introduction}

As a natural phenomenon of non-thermal filamentary discharges with a large amount of applications, a streamer discharge happens when an insulating medium such as air is exposed to a sufficiently strong electric field, where electron avalanche occurs and forms filamentary streamers. The filamentary streamer discharges are pivotal for many gas discharges in nature \cite{ebert2008streamers,jap}, e.g., the lightning \cite{sadighi2015streamer} and the sprite discharges in high altitude \cite{liu2015sprite,moudry2003imaging}. It also has mature industrial applications \cite{bogaerts2002gas, babaeva2006streamer, samaranayake2000pulsed} like dust precipitator, ozone production, and water purification \cite{vsunka2001pulse, joshi2013streamer}. A review can be found in \cite{ebert2006multiscale}.

Streamers can be classified into positive and negative ones. The photoionization plays an important role in the propagation of streamers in air, especially for positive ones. In particular, the photoionization provides seed electrons ahead of the tips, which are required by the propagation of positive streamers \cite{won2002experimental, kulikovsky2000, pancheshnyi2001role, ldg, weno}. Besides, the stochastic photoionization is shown to have an impact on the branching of streamers \cite{xiong2014branching, bagheri2019effect, marskar20203d}. 

Due to the importance of the photoionization to streamer discharges, its modelling and simulation have attracted continuous attention. The classical model for oxygen-nitrogen mixture derived by Zheleznyak \textit{et al.} in \cite{zhelezniak1982} is widely utilized in the simulation of positive streamers \cite{pancheshnyi2005role, stephens2018practical}, and was improved in \cite{pancheshnyi2014, jiang2018photoionization} to gain better accuracy and has been extended to a stochastic version in \cite{chanrion2008}.

Direct calculation of the classical integral model requires a large amount of computation, especially in three dimensions (3D) where streamer discharges inherently happen. To ease the numerical difficulty and reduce the computational cost, some approximation methods were proposed in \cite{bourdon2007, liu2007apl, luque2007} based on the kernel expansion and conversion to Helmholtz equations. Moreover, modeling of photoionization based on the radiative transfer equation (RTE) also provides good results \cite{fvmrte2008}.

Less than two decades ago, the kernel-independent fast multipole method (FMM) was proposed to compute particle interactions efficiently and accurately \cite{ying2004}. It can be easily applied to the convolutional integrals \cite{malhotra2015pvfmm}, and its computational complexity is comparable to the method of fast Fourier transform (FFT). Compared with FFT, FMM can be applied to more general computational domains and has better parallel efficiency in distributed computations. In addition, it can be directly applied on a broad class of different integral forms compared with the kernel-dependent FMM which requires specific kernel expansion and efficient translation for different  kernels \cite{greengard1987fast}.

Motivated by the good performance of the kernel-independent FMM, this paper extends its application to the computation of photoionization rates, and focuses on the following properties: (i) accuracy and robustness for different pressures, (ii) good efficiency, and (iii) extensibility to other integral models. The rest of this paper is organized as follows. The classical integral method and its associated PDE-based approximations are reviewed in Section \ref{model}. Section \ref{fmm} introduces the fast multipole method on a general numerical integral form. The quantified performance of the fast multipole method and comparisons with other approximations for computing photoionization are presented in Section \ref{comparison}, and for computing
streamer discharges are reported in Section \ref{comparison1}. Finally, conclusions are drawn in Section \ref{conclusion}.

%In Ref.~\cite{ref1}...
%In Refs.~\cite{ref1,ref2}...
%On webpage~\cite{web}...

\section{Model formulation}\label{model}

To make the contents self-contained, we briefly review commonly used approaches for photoionization calculations. \addi{We focus on deterministic or continuum models in this paper, and readers interested in stochastic models using Monte Carlo collision method with discrete photon particles could refer to \cite{bagheri2019effect,marskar20203d,chanrion2008}.}

\subsection{Classical integral photoionization model by Zheleznyak \textit{et al.}}\label{classint}

The widely used photoionization model derived by Zheleznyak \textit{et al.} \cite{zhelezniak1982} describes the photoionization rate by
\begin{equation}
S_{\rm ph}(\vec{x}) = \iiint_{V'} \frac{I(\vec{y})g(|\vec{x}-\vec{y}|)}{4\pi |\vec{x}-\vec{y}|^2}\mathrm{d}\vec{y}, \qquad \forall \vec{x} \in V,
\label{integral}
\end{equation}
where $\vec{x} = (x, y, z)^T$, $V'$ is the source chamber in which the photons are emitted, and $V$ is the collector chamber where the photons are absorbed, $I(\vec{y})$ is proportional to the intensity of the source radiation:
\begin{equation}
I(\vec{y}) = \xi \frac{p_q}{p+p_q} \frac{\omega}{\alpha} S_i(\vec{y}),
\label{Ifun}
\end{equation}
where $\xi$ is the photoionization efficiency, $p_q$ is the quenching pressure, $p$ is the gas pressure, $\omega$ and $\alpha$ are the excitation coefficient of emitting states without quenching processes and the effective Townsend ionization coefficient, respectively, with $\frac{\omega}{\alpha}$ being a coefficient to be determined by experiments, and $S_i$ is the effective ionization rate. The function $g(r)=g(|\vec{x}-\vec{y}|)$ in (\ref{integral}) is given by
\begin{equation}
\frac{g(r)}{p_{_{O_2}}} = \frac{\exp(-\chi_{\min}\, p_{_{O_2}}r) - \exp(-\chi_{\max}\,p_{_{O_2}}r)}{p_{_{O_2}}r \ln(\chi_{\max}/\chi_{\min})},
\label{gfun}
\end{equation}
where $r=|\vec{x}-\vec{y}|$, $p_{_{O_2}}$ is the partial pressure of oxygen, $\chi_{\max}=2$\,cm$^{-1}$\,Torr$^{-1}$ and $\chi_{\min}=0.035$\,cm$^{-1}$\,Torr$^{-1}$ are the maximum and minimum absorption coefficients of O$_2$ in wavelength 980-1025\,\AA, respectively, as indicated in \cite{zhelezniak1982}. Note that when we define $g(r)$ in (\ref{gfun}), we follow \cite{zhelezniak1982, pancheshnyi2014,bourdon2007} to write $g(r)/p_{_{O_2}}$ on the left-hand side so that the right-hand side is dependent directly on the product $p_{_{O_2}}r$. Interested readers may refer to \cite{zhelezniak1982} and \cite{pancheshnyi2014} for more details.

Clearly, Eq. (\ref{integral}) is a convolution in 3D. A naive numerical implementation of (\ref{integral}) requires a whole domain quadrature for every point $\vec{x} \in V$, which requires a time complexity $O(N^2)$ with $N$ being the total number of degrees of freedom. One idea to reduce the computational cost is to use a coarse grid in the weak field at the price of possibly losing some accuracy, as \cite{kulikovsky2000} did in the 3D cases with cylindrical symmetry. 

\subsection{Exponential or Helmholtz PDE approximation}\label{Helmholtz}
Instead of a straightforward computation of the integral, the efficiency can be significantly enhanced by converting it into a problem of differential equations at the expense of losing some accuracy. One important and pioneer work was done in \cite{luque2007}, which approximates the photoionization kernel as the sum of the fundamental solutions of a number of partial differential equations. The function $g(r)$ defined by (\ref{gfun}) is approximated as follows:
\begin{equation}
\frac{g(r)}{p_{_{O_2}}} \approx p_{_{O_2}} r \sum_{j=1}^{N_{E}} C_j \exp(-\lambda_j p_{_{O_2}} r),
\label{gexp}
\end{equation}
where $\lambda_j$ and $C_j$ ($1 \leq j \leq N_E$) are constants that can be fit numerically \cite{bourdon2007,luque2007}. Consequently, it suffices to take the linear combination of $S_{{\rm ph},j}$ to approximate the integral (\ref{integral})
\begin{equation}
S_{\rm ph}(\vec{x}) \approx \sum_{j=1}^{N_{E}}C_j S_{{\rm ph},j}(\vec{x}),
\label{sphHel}
\end{equation}
where $S_{{\rm ph},j}(\vec{x})$ is the solution of the following modified Helmholtz equation
\begin{equation}
(-\Delta + (\lambda_j p_{_{O_2}})^2 ) S_{{\rm ph},j}(\vec{x}) = (p_{_{O_2}})^2I(\vec{x}).
\label{Hel}
\end{equation}

The modified Helmholtz equation \eqref{Hel} can be solved efficiently by numerous fast elliptic solvers like multigrid-preconditioned FGMRES method \cite{lin2018}.

$N_E=2$ was used in \cite{luque2007}, and the constants $\lambda_j$ and $C_j$ were chosen to fit the low-pressure experimental data from \cite{penney1970} (the misprint of these constants is corrected in \cite{dubinova2016}). $N_E=3$ was suggested in \cite{bourdon2007} for a better fitting for the range $1<p_{_{O_2}}r<150$\,Torr$\cdot$cm, and the constants $\lambda_j$ and $C_j$ are chosen to fit the function in Eq. (\ref{gexp}) since it agrees well with the experimental data in both low-pressure and atmospheric airs \cite{penney1970, aints2002}, as indicated in \cite{naidis2006}. While zero boundary conditions were used in \cite{luque2007}, it is suggested in \cite{bourdon2007} that the boundary condition for Eq. (\ref{Hel}) can be provided by computing the integral (\ref{integral}). In this paper, we take the three-term exponential approximation and adopt the coefficients in \cite{bourdon2007} listed in Table \ref{Helcoeff}.

\begin{table}
\caption{Coefficients of three-exponential ($N_{E}=3$) approximation in (\ref{gexp}) \cite{bourdon2007}.}
\medskip
\centering
\begin{tabular}{ccc}
\hline
$j$ & $C_j$ (cm$^{-2}$\,Torr$^{-2}$)& $\lambda_j $ (cm$^{-1}$\,Torr$^{-1}$) \\
\hline
1 & $1.986 \times 10^{-4}$ & $0.0553$ \\
2 & $0.0051$ & $0.1460$ \\
3 & $0.4886$ & $0.89$ \\
\hline
\end{tabular}\label{Helcoeff}
\end{table}

\subsection{Three-group radiative transfer approximation}\label{Threegroup}
Another type of differential equations that can facilitate the computation of the photoionization rate is the radiative transfer equation. In \cite{bourdon2007,fvmrte2008,segur2006}, the following multi-group approximation of the steady-state radiative transfer equation is chosen to describe the intensity of radiation $\Psi_j$ for the $j$-th group of spectral frequency:
\begin{equation}
\vec{\omega} \cdot \nabla \Psi_j(\vec{x}, \vec{\omega}) + \kappa_j \Psi_j(\vec{x}, \vec{\omega}) = \frac{n_{u}(\vec{x})}{4 \pi c\, \tau_u}, \quad j = 0,1,\cdots,N_{\nu},
\label{rte}
\end{equation}
where $\vec{\omega} \in S^2$ is the solid angle defined on the unit sphere, $\kappa_j$ is the absorption coefficient, $n_u$ is the density of the species with the excited state $u$, $c$ is the speed of light and $\tau_u$ is the radiative relaxation time for the state $u$. Here the scattering and the change in frequency of the photons during collisions with molecules have been neglected \cite{fvmrte2008,segur2006}. For photoionization in air, $\kappa_j = \lambda_j\, p_{_{O_2}}$, and for simplicity, only one excited state is considered
\begin{equation}
\frac{n_u(\vec{x})}{\tau_u} = \frac{I(\vec{x})}{\xi},
\end{equation}
with $\lambda_j$ to be determined by data fitting \cite{zhelezniak1982,bourdon2007,segur2006}. The photoionization rate is then proportional to the weighted sum of the integral of $\Psi_j$ over $\vec{\omega} \in S^2$:
\begin{equation}
\begin{split}
S_{\rm ph}(\vec{x}) &= \sum_{j=1}^{N_{\nu}} A_j \,\xi\, p_{_{O_2}}c \int_{S^2} \Psi_j(\vec{x},\vec{\omega})
\mathrm{d}\vec{\omega}, \\
&=\sum_{j=1}^{N_{\nu}} A_j\, \xi\, p_{_{O_2}}c \iiint _{V} \frac{n_u(\vec{y})}{c \, \tau_u} \frac{\exp(-\lambda_j p_{_{O_2}}|\vec{x}-\vec{y}|)}{4 \pi |\vec{x}-\vec{y}|^2} \mathrm{d}\vec{y},
\end{split}
\label{rteSph}
\end{equation}
where $A_j$ are also parameters which can be fit according to the experimental data.
To determine the parameters, it is noticed that \eqref{rteSph} is identical to (\ref{integral}) if
\begin{equation}
\sum_{j=1}^{N_{\nu}} A_j p_{_{O_2}} \exp(-\lambda_j p_{_{O_2}} r) = g(r), \qquad r=|\vec{x}-\vec{y}|,
\label{rtetoint}
\end{equation}
where $g(r)$ is given in (\ref{gfun}), and the coefficient $A_j$ and $\lambda_j$ ($1 \leq j \leq N_{\nu}$) are determined by fitting the left hand side of (\ref{rtetoint}) with $g(r)$ in the range $0.1 < p_{_{O_2}}r < 150$\,Torr$\cdot$cm \cite{bourdon2007}. The results for three-group ($N_{\nu}=3$) approximation are shown in Table \ref{threegroupcoeff}.

\begin{table}
\caption{Coefficients of three-group ($N_{\nu}=3$) approximation in (\ref{rtetoint}) \cite{bourdon2007}.}
\medskip
\centering
\begin{tabular}{ccc}
\hline
$j$ & $A_j$ (cm$^{-1}$\,Torr$^{-1}$)& $\lambda_j$ (cm$^{-1}$\,Torr$^{-1}$) \\
\hline
1 & 0.0067 & 0.0447 \\
2 & 0.0346 & 0.1121 \\
3 & 0.3059 & 0.5994 \\
\hline
\end{tabular}\label{threegroupcoeff}
\end{table}

Instead of computing the integral in \eqref{rteSph}, a more efficient way to get the intensity function $\Psi_j$ is to solve (\ref{rte}) as an differential equation. For example, in \cite{fvmrte2008}, a direct solver of (\ref{rte}) was employed for two-dimensional axisymmetric discharges using the finite volume method for both space and angular variables.

However, the radiative transfer equation \eqref{rte} is still a five-dimensional partial differential equation. Further reduction of dimensionality can be realized by the improved Eddington or SP$_3$ approximation \cite{bourdon2007, segur2006, tmag2020}. In \cite{larsen2002}, the simplified P$_N$ (SP$_N$) approximations of optically thick radiative heat transfer equations are theoretically derived by asymptotic analysis. SP$_N$ approximations are introduced in \cite{segur2006} to obtain a fast numerical simulation for the photoionization source term mainly with monochromatic (one-group) approximation. The SP$_N$ approximations for photoionization are further improved in \cite{bourdon2007}, and extended to multi-group approximation, including the three-group SP$_3$ method which approximates the isotropic part of the solution by \cite{ bourdon2007,segur2006}
\begin{equation}
\int_{S^2} \Psi_j(\vec{x},\vec{\omega}) \mathrm{d}\vec{\omega} = \frac{\gamma_2 \phi_{j,1} (\vec{x}) - \gamma_1 \phi_{j,2} (\vec{x})}{\gamma_2-\gamma_1},
\label{sp3sol}
\end{equation}
where $\gamma_n = \frac{5}{7}\left[ 1 + (-1)^n 3 \sqrt{\frac{6}{5}}\right]$ with $n=1,2$, and $\phi_{j,1} (\vec{x})$ and $\phi_{j,2} (\vec{x})$ are solutions of the following two Helmholtz equations
\begin{align}
\left(- \Delta + \frac{(\lambda_j p_{_{O_2}})^2}{\mu_1^2} \right) \phi_{j,1}(\vec{x}) = \frac{\lambda_j p_{_{O_2}}}{\mu_1^2} \frac{n_u(\vec{x})}{c\,\tau_u}, \label{phi1} \\
\left(- \Delta + \frac{(\lambda_j p_{_{O_2}})^2}{\mu_2^2} \right) \phi_{j,2}(\vec{x}) = \frac{\lambda_j p_{_{O_2}}}{\mu_2^2} \frac{n_u(\vec{x})}{c\,\tau_u} , \label{phi2}
\end{align}
with the coefficients $\mu_{n}=\sqrt{\frac{3}{7} + (-1)^n \frac{2}{7}\sqrt{\frac{6}{5}}}$ ($n=1,2$). The equations (\ref{phi1})-(\ref{phi2}) need to be equipped with proper boundary conditions (BCs). In \cite{bourdon2007}, the BCs are obtained directly from the integral model (\ref{integral}), which requires numerical integrations over the whole domain for all the grid points on the boundary. Later in \cite{liu2007apl}, the same authors proposed the following more efficient BCs based on \cite{larsen2002} for a boundary surface without reflection and emission:
\begin{align}
\nabla \phi_{j,1} \cdot \vec{n} + \alpha_1 (\lambda_j p_{_{O_2}}) \phi_{j,1} =- \beta_2 (\lambda_j p_{_{O_2}}) \phi_{j,2}, \label{boundary1} \\
\nabla \phi_{j,2} \cdot \vec{n} + \alpha_2 (\lambda_j p_{_{O_2}}) \phi_{j,2} =- \beta_1 (\lambda_j p_{_{O_2}}) \phi_{j,1}, \label{boundary2}
\end{align}
where $\vec{n}$ is the outward unit normal vector, $\alpha_n = \frac{5}{96} \left( 34 + (-1)^{n-1}11 \sqrt{\frac{6}{5}} \right)$ and $\beta_n = \frac{5}{96} \left( 2 + (-1)^n \sqrt{\frac{6}{5}} \right)$ ($n=1,2$).

\section{Fast multipole method for accurate and efficient evaluation of integral}\label{fmm}
As can be seen from Sections \ref{Helmholtz} and \ref{Threegroup}, different methods based on differential equations have been proposed to approximate the integral (\ref{integral}) or (\ref{rteSph}), leading to much higher numerical efficiency than directly computing the integral (\ref{integral}). However, the approximation errors of these methods might be significant in some cases. On the other hand, despite the high computational cost \cite{dubinova2016}, the results calculated from the integral form are free of further approximations, therefore, these results are often used as reference solutions \cite{bourdon2007, segur2006,celestin2008}. Moreover, the integral form can be easily extended to stochastic versions \cite{chanrion2008, teunissen2016}. The importance of the integral form inspires us to tackle the original integration problem (\ref{integral}) directly using fast algorithms. The exponential decay of the kernel with respect to the distance (see \eqref{gfun}) reminds us to adopt the efficient and accurate fast multipole method \cite{ying2004,greengard1987fast}, which utilizes the low-rank structure of far-away interactions to gain significant speed-up.

The fast multipole method used in this paper {\color{black}\cite{ying2004}} is established based on the fast evaluation of the numerical quadrature of (\ref{integral}). For convenience, we discretize $S_{\rm ph}$ and $n_e$ on the same mesh. In general, the integral (\ref{integral}) can be discretized as
\begin{equation}
S_{\rm ph}(\vec{x}_i) = \sum_{j=1}^{N_{\rm pt}} G(\vec{x}_i ,\vec{y}_j) I(\vec{y}_j), \qquad i=1,\cdots,N_{\rm pt},
\label{summation}
\end{equation}
where $G(\cdot,\cdot)$ is the discrete kernel function calculated from the corresponding function in (\ref{integral}) and the numerical quadrature weights. In this paper, we apply the midpoint quadrature rule on each grid cell unless $\vec{x}_i$ and $\vec{y}_j$ are in the same grid where the second-order Gauss-Legendre quadrature rule is alternatively applied. \addii{More specifically, the points $\vec{y}_j$ in \eqref{summation} are taken as the centers of each cell in the given mesh. As a result, $N_{\rm pt}$ is the number of cells in the mesh, and $I(\vec{y}_j)$ could be evaluated from \eqref{Ifun} where $S_i(\vec{y}_j)$ is calculated locally by discrete values at this cell. In practical implementation, we further multiply $I(\vec{y}_j)$ by the volume of cell (quadrature weight) at $\vec{y}_j$ for all $N_{\rm pt}$ points. If a uniform mesh is applied for discretization, the multiplication factor is $h_x h_y h_z$ where $h_x$, $h_y$ and $h_z$ are mesh size in $x$, $y$ and $z$, respectively. With this multiplication, $G(\cdot,\cdot)$ does not contain factor $h_x h_y h_z$ and can be written as
\[
G(\vec{x}_i, \vec{y}_j) = \left\{ 
\begin{array}{ll}
\frac{g(\vec{x}_i - \vec{y}_j)}{4 \pi |\vec{x}_i - \vec{y}_j|^2}, & i \neq j, \\
\frac{3g(\sqrt{(h_x^2 + h_y^2 + h_z^2)/12})}{\pi (h_x^2 + h_y^2 + h_z^2)}, & i = j,
\end{array}
\right.
\]
where $g(\cdot)$ comes from \eqref{gfun}.} We remark that this is not essential and other numerical quadrature can also be used.

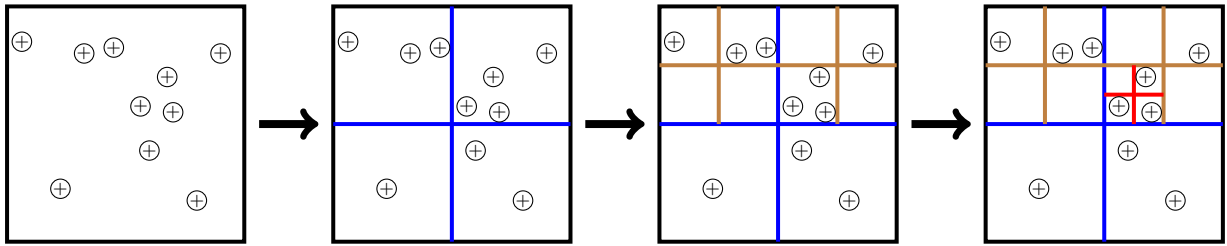
\begin{figure}[htb!]
\centering
\begin{tikzpicture}[scale=0.78]
\draw[line width=1.5pt] (0,0) rectangle (4,4);
\draw[->, line width=3pt] (4.25,2) -- (5.25,2);
\draw[line width=1.5pt] (5.5,0) rectangle (9.5,4);
\draw[->, line width=3pt] (9.75,2) -- (10.75,2);
\draw[line width=1.5pt] (11,0) rectangle (15,4);
\draw[->, line width=3pt] (15.25,2) -- (16.25,2);
\draw[line width=1.5pt] (16.5,0) rectangle (20.5,4);
\node[draw=black, circle, inner sep =0pt] at (0.9,0.9) {\tiny{+}};
\node[draw=black, circle, inner sep =0pt] at (3.2,0.7) {\tiny{+}};
\node[draw=black, circle, inner sep =0pt] at (2.4,1.55) {\tiny{+}};
\node[draw=black, circle, inner sep =0pt] at (2.25,2.3) {\tiny{+}};
\node[draw=black, circle, inner sep =0pt] at (0.25,3.4) {\tiny{+}};
\node[draw=black, circle, inner sep =0pt] at (3.6,3.2) {\tiny{+}};
\node[draw=black, circle, inner sep =0pt] at (2.7,2.8) {\tiny{+}};
\node[draw=black, circle, inner sep =0pt] at (2.8,2.2) {\tiny{+}};
\node[draw=black, circle, inner sep =0pt] at (1.8,3.3) {\tiny{+}};
\node[draw=black, circle, inner sep =0pt] at (1.3,3.2) {\tiny{+}};
\draw[blue, line width=1.5pt] (5.5,2) -- (9.5,2);
\draw[blue, line width=1.5pt] (7.5,0) -- (7.5,4);
\node[draw=black, circle, inner sep =0pt] at (6.4,0.9) {\tiny{+}};
\node[draw=black, circle, inner sep =0pt] at (8.7,0.7) {\tiny{+}};
\node[draw=black, circle, inner sep =0pt] at (7.9,1.55) {\tiny{+}};
\node[draw=black, circle, inner sep =0pt] at (7.75,2.3) {\tiny{+}};
\node[draw=black, circle, inner sep =0pt] at (5.75,3.4) {\tiny{+}};
\node[draw=black, circle, inner sep =0pt] at (9.1,3.2) {\tiny{+}};
\node[draw=black, circle, inner sep =0pt] at (8.2,2.8) {\tiny{+}};
\node[draw=black, circle, inner sep =0pt] at (8.3,2.2) {\tiny{+}};
\node[draw=black, circle, inner sep =0pt] at (7.3,3.3) {\tiny{+}};
\node[draw=black, circle, inner sep =0pt] at (6.8,3.2) {\tiny{+}};
\draw[blue, line width=1.5pt] (11,2) -- (15,2);
\draw[blue, line width=1.5pt] (13,0) -- (13,4);
\draw[brown, line width=1.5pt] (11,3) -- (15,3);
\draw[brown, line width=1.5pt] (12,2) -- (12,4);
\draw[brown, line width=1.5pt] (14,2) -- (14,4);
\node[draw=black, circle, inner sep =0pt] at (11.9,0.9) {\tiny{+}};
\node[draw=black, circle, inner sep =0pt] at (14.2,0.7) {\tiny{+}};
\node[draw=black, circle, inner sep =0pt] at (13.4,1.55) {\tiny{+}};
\node[draw=black, circle, inner sep =0pt] at (13.25,2.3) {\tiny{+}};
\node[draw=black, circle, inner sep =0pt] at (11.25,3.4) {\tiny{+}};
\node[draw=black, circle, inner sep =0pt] at (14.6,3.2) {\tiny{+}};
\node[draw=black, circle, inner sep =0pt] at (13.7,2.8) {\tiny{+}};
\node[draw=black, circle, inner sep =0pt] at (13.8,2.2) {\tiny{+}};
\node[draw=black, circle, inner sep =0pt] at (12.8,3.3) {\tiny{+}};
\node[draw=black, circle, inner sep =0pt] at (12.3,3.2) {\tiny{+}};
%------------------------------
\draw[blue, line width=1.5pt] (16.5,2) -- (20.5,2);
\draw[blue, line width=1.5pt] (18.5,0) -- (18.5,4);
\draw[brown, line width=1.5pt] (16.5,3) -- (20.5,3);
\draw[brown, line width=1.5pt] (17.5,2) -- (17.5,4);
\draw[brown, line width=1.5pt] (19.5,2) -- (19.5,4);
\draw[red, line width=1.5pt] (18.5,2.5) -- (19.5,2.5);
\draw[red, line width=1.5pt] (19,2) -- (19,3);
\node[draw=black, circle, inner sep =0pt] at (17.4,0.9) {\tiny{+}};
\node[draw=black, circle, inner sep =0pt] at (19.7,0.7) {\tiny{+}};
\node[draw=black, circle, inner sep =0pt] at (18.9,1.55) {\tiny{+}};
\node[draw=black, circle, inner sep =0pt] at (18.75,2.3) {\tiny{+}};
\node[draw=black, circle, inner sep =0pt] at (16.75,3.4) {\tiny{+}};
\node[draw=black, circle, inner sep =0pt] at (20.1,3.2) {\tiny{+}};
\node[draw=black, circle, inner sep =0pt] at (19.2,2.8) {\tiny{+}};
\node[draw=black, circle, inner sep =0pt] at (19.3,2.2) {\tiny{+}};
\node[draw=black, circle, inner sep =0pt] at (18.3,3.3) {\tiny{+}};
\node[draw=black, circle, inner sep =0pt] at (17.8,3.2) {\tiny{+}};
\end{tikzpicture}
\caption{\label{htree}An example of a hierarchical tree in 2D, with $N_{\rm pt}=10$, $m=2$. The arrows show the construction procedure, and the circles with ``$+$'' denote the $N_{\rm pt}$ points.}
\end{figure}

The kernel-independent adaptive fast multipole method \cite{ying2004} does not require the implementation of multipole expansions \cite{greengard1987fast, cheng1999fast} of the kernel function. Based on a hierarchical tree, it uses a continuous equivalent density on a surface enclosing a box to represent the potential generated by sources inside the box. Given a set of $N_{\rm pt}$ points in three dimensions, a hierarchical octree is constructed adaptively such that each leaf cube of the tree contains no more than $m$ points, where $m$ is a selected constant. This octree can be built from a sufficiently large root cube to contain all $N_{\rm pt}$ points, and then subdivided to equal-sized sub-cubes recursively if the current cube contains more than $m$ points. For illustrative purpose, an example of the hierarchical tree in two dimensions (2D), i.e. quadtree, is shown in Figure \ref{htree}.

\begin{figure}[btp!]
\centering
\begin{tikzpicture}[scale=0.8]
\fill[red!20!white] (0,0) rectangle (6,6);
\fill[green!20!white] (2,1) rectangle (5,4);
\draw (0,0) grid (6,6);
\draw[blue, line width=1.5pt] (3,2) rectangle (4,3);
\node[draw=green!20!white, circle, inner sep =0pt] at (3.5,2.5) {$B$};
\node[draw=red!20!white, rectangle, inner sep =0pt] at (1.1,5.5) {$\mathcal{F}(B)$};
\node[draw=green!20!white, rectangle, inner sep =0pt] at (3,3.5) {$\mathcal{N}(B)$};
\end{tikzpicture}
\caption{\label{fNF}Cross section of near range $\mathcal{N}(B)$ and far range $\mathcal{F}(B)$ of a box $B$ in 3D. The blue thick side is the boundary of $B$, green part is $\mathcal{N}(B)$ and red part is $\mathcal{F}(B)$.}
\end{figure}
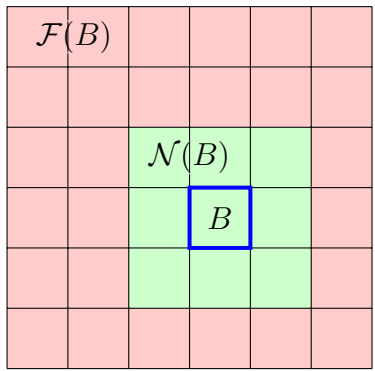

To sketch the idea, we consider the simple case where the source points are uniformly distributed. This corresponds to the case when the uniform mesh is applied in the discretization of $I(\vec{y})$ in (\ref{Ifun}). In this case, for each target point $\vec{x}_{i}$ in a cube or box $B$, fast multipole method splits the summation (\ref{summation}) into two parts, namely, near interactions and far interactions:
\begin{equation}
S_{\rm ph}(\vec{x}_i) = \sum_{\vec{y}_j \in \mathcal{N}(B)} G(\vec{x}_i ,\vec{y}_j) I(\vec{y}_j) + \sum_{\vec{y}_j \in \mathcal{F}(B)} G(\vec{x}_i ,\vec{y}_j) I(\vec{y}_j),
\label{nearfar}
\end{equation}
where $\mathcal{N}(B)$ and $\mathcal{F}(B)$ are the near range and far range of $B$, respectively. For $\vec{y}_j \in \mathcal{N}(B)$, the interactions with all $\vec{x}_i \in B$ are calculated directly. For the points $\vec{y}_j \in \mathcal{F}(B)$, the interactions can be approximated with controlled accuracy due to the low-rankness of $G(\vec{x}_i ,\vec{y}_j)$. If a box is centered at $\vec{c}$ with side length $2r$, then $\mathcal{N}(B)$ is defined as a box centered at $\vec{c}$ with side length $6r$, and $\mathcal{F}(B)$ is the domain outside $\mathcal{N}(B)$ (See Figure \ref{fNF}).

In (\ref{nearfar}), the summation for points $\vec{y}_j \in \mathcal{F}(B)$ can be approximated using the hierarchy tree. The idea is composed of two parts: 1) represent the potential generated from source points inside any box $B$ by some equivalent source points enclosing $B$; 2) represent the potential generated from source points in $\mathcal{F}(B)$ by other equivalent source points enclosing $B$, which gives an approximation to the summation for points $\vec{y}_j \in \mathcal{F}(B)$ in (\ref{nearfar}). The first part is implemented by post-order traversal of the hierarchical tree. If $B$ is a leaf box, the potential generated from the source points inside $B$ is represented by several equivalent points surrounding the box, as is called the multipole expansion to be defined in (\ref{s2m}). If $B$ is not a leaf box, its multipole expansion can be accumulated from the multipole expansion of all its children boxes by ``M2M translation'' to be defined in (\ref{m2m}). With the help of the equivalent source points in the first part, we can approximate the potential in $B$ from original source points in $\mathcal{F}(B)$ by a small number of equivalent source points in $\mathcal{F}(B)$ calculated from the first part. This is the idea of the second part, and we similarly represent the potential generated from source points in $\mathcal{F}(B)$ by some equivalent source points surrounding $B$, as is called the local expansion to be defined in (\ref{s2l}). The second part is implemented by pre-order traversal of the hierarchical tree. If a non-root box $B$ is embedded in its parent box $\mathcal{P}(B)$, its local expansion is calculated from: the accumulation of the local expansion of $\mathcal{P}(B)$, which is called ``L2L translation'' to be defined in (\ref{l2l}); and the multipole expansion of the boxes in $\mathcal{N}(\mathcal{P}(B))$ but not adjacent to $B$, as it is implemented by the operation called ``M2L translation'' to be defined (\ref{m2l}).

\begin{figure}[htb!]
\centering
\begin{tikzpicture}
\fill[green!20!white] (-2.4,-2.4) rectangle (2.4,2.4);
\draw[line width = 1pt] (-0.8,-0.8) rectangle (0.8,0.8);
\node[draw=green!20!white, circle, inner sep =0pt] at (-0.4,0.4) {$B$};
\draw[dashed, line width=1pt] (-0.96,-0.96) rectangle (0.96,0.96);
\draw[densely dotted, line width = 1pt] (-2.08,-2.08) rectangle (2.08,2.08);
\draw [red, line width=4pt,line cap=round,dash pattern=on 0pt off 0.64cm] (-0.96,-0.96) rectangle (0.96,0.96);
\draw [blue, line width=4pt,line cap=round,dash pattern=on 0pt off 1.38667cm] (-2.08,-2.08) rectangle (2.08,2.08);
\node[draw=black, circle, inner sep =0pt, thick] at (0.1,-0.1) {\tiny{\textbf{+}}};
\node[draw=black, circle, inner sep =0pt, thick] at (-0.2,-0.1) {\tiny{\textbf{+}}};
\node[draw=black, circle, inner sep =0pt, thick] at (0.05,0.2) {\tiny{\textbf{+}}};
\node[circle, inner sep =0pt, text=blue] at (-1.5,1.7) {$x^{B,u}$};
\node[circle, inner sep =0pt, text=red] at (-1,-1.3) {$y^{B,u}$};
\draw[->, line width = 1pt, blue] (0,0.4) -- (0,2.08) node[midway, above left] {{\footnotesize $(1)$}};
\draw[->, line width = 1pt, red] (2.08,0) -- (0.96,0) node[midway, below] {{\footnotesize $(2)$}};
\fill[green!20!white] (4.6,-2.4) rectangle (9.4,2.4);
\draw[line width = 1pt] (6.2,-0.8) rectangle (7.8,0.8);
\node[draw=green!20!white, circle, inner sep =0pt] at (6.6,0.4) {$B$};
\draw[densely dotted, line width=1pt] (6.04,-0.96) rectangle (7.96,0.96);
\draw[dashed, line width = 1pt] (4.92,-2.08) rectangle (9.08,2.08);
\draw [blue, line width=4pt,line cap=round,dash pattern=on 0pt off 0.64cm] (6.04,-0.96) rectangle (7.96,0.96);
\draw [red, line width=4pt,line cap=round,dash pattern=on 0pt off 1.38667cm] (4.92,-2.08) rectangle (9.08,2.08);
\node[draw=black, circle, inner sep =0pt, thick] at (10.1,-0.1) {\tiny{\textbf{+}}};
\node[draw=black, circle, inner sep =0pt, thick] at (9.8,-0.1) {\tiny{\textbf{+}}};
\node[draw=black, circle, inner sep =0pt, thick] at (10.05,0.2) {\tiny{\textbf{+}}};
\node[circle, inner sep =0pt, text=red] at (5.5,1.7) {$y^{B,d}$};
\node[circle, inner sep =0pt, text=blue] at (6,-1.3) {$x^{B,d}$};
\draw[->, line width = 1pt, blue] (9.7,0) -- (7.96,0) node[midway, above left] {{\footnotesize $(1)$}};
\draw[->, line width = 1pt, red] (7,0.96) -- (7,2.08) node[midway, right] {{\footnotesize $(2)$}};
\end{tikzpicture}
\caption{\label{fmulloc}Cross section of equivalent surfaces and check surfaces in multipole expansion (left subfigure) and local expansion (right subfigure) of box $B$. Dashed lines with red dots denote equivalent surfaces, where red dots can be viewed as equivalent sources. Dotted lines with blue dots denote check surface, where blue dots can be viewed as check points. Green shadow is the near range of $B$. Circles with ``+'' denote source points. Step $(1)$ in blue arrow is the evaluation of potential on check surface, and step $(2)$ in red arrow is the calculation of equivalent density on equivalent surface.}
\end{figure}
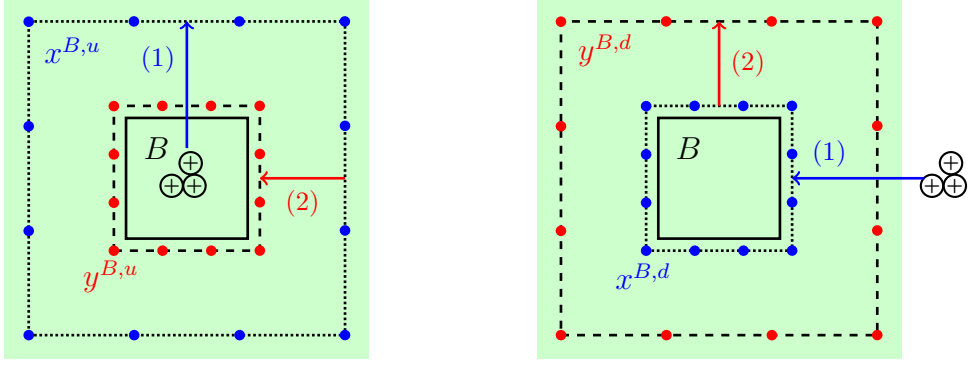

We now show more details about the kernel-independent FMM: firstly introduce multipole expansion and local expansion in the FMM, and then show three translations among them: M2M (multipole to multipole), M2L (multipole to local) and L2L (local to local). For simplicity, we would like to neglect the vector symbol on $\vec{x}$ and $\vec{y}$ when introducing FMM.
\paragraph{Multipole expansion}
Multipole expansion of a box $B$ is used to represent the potential in $\mathcal{F}(B)$, generated by the source inside $B$. Two surfaces of the cube are introduced for the approximation, upward equivalent surface $y^{B,u}$ and upward check surface $x^{B,u}$. The equivalent surface $y^{B,u}$ should be taken to enclose $B$, and check surface $x^{B,u}$ encloses equivalent surface $y^{B,u}$. Moreover, both $y^{B,u}$ and $x^{B,u}$ should locate inside $\mathcal{N}(B)$. See these two box surfaces in Figure \ref{fmulloc}.

An upward density function $\phi^{B,u}(y)$, or the density $\phi^{B,u}_k=\phi^{B,u}(y_k^{B,u})$ on several upward equivalent source points $y_k^{B,u} \in y^{B,u}$, is introduced to represent the potential in $\mathcal{F}(B)$, generated by the source inside $B$. If the upward check potential $q^{B,u}(x)$ at the check surface $x^{B,u}$, evaluated from the source in $B$, is equal to the potential $q^{B,u}(x)$ evaluated from the equivalent source $\phi^{B,u}_k$, then these source density points $\phi^{B,u}_k$ can be used to represent the potential outside the check surface $x^{B,u}$ including $\mathcal{F}(B)$. This is because of the uniqueness of the Dirichlet boundary value problem (similar to the method of image charges in electrostatics). The equality is written as
\begin{equation}
\sum_{k \in Id(y^{B,u})} G(x_j^{B,u},y_k^{B,u}) \phi_k^{B,u}= q^{B,u}(x_j^{B,u}) = \sum_{i \in I^{B}_s} G(x_j^{B,u},y_i) I(y_i),\ \forall j \in Id(x^{B,u}),
\label{s2m}
\end{equation}
where $I_s^B$ is the index set of the source points inside $B$, $Id(y^{B,u})$ is the index set of discrete source points on $y^{B,u}$ and $Id(x^{B,u})$ is the index set of discrete check points on $x^{B,u}$. A prescribed number $m_0$ is used to denote the number of discrete equivalent source points at each side of $y^{B,u}$, and this number is identical to the number of check points at each side of $x^{B,u}$. \addii{Equation \eqref{s2m} is solved by calculating the upward check potential $q^{B,u}(x_j^{B,u})$ from the second equality, and then solving a linear system to get the upward equivalent density $\phi_k^{B,u}$ from $q^{B,u}(x_j^{B,u})$. This two-step procedure is also applicable to the local expansion \eqref{s2l}. For illustration, we marked the first step as blue arrow and second step as red arrow in Figure \ref{fmulloc}.}
\paragraph{Local expansion} Local expansion is used to represent the potential inside a box $B$, generated by the source in $\mathcal{F}(B)$. Similar to the multipole expansion, a downward equivalent surface $y^{B,d}$ with downward equivalent density $\phi^{B,d}$ on it, is introduced. At the same time, downward check surface $x^{B,d}$ with downward check potential $q^{B,d}$ is used to check the equality of potential generated by the source in $\mathcal{F}(B)$ and the one generated by $\phi^{B,d}$. Different from the multipole expansion, $y^{B,d}$ should enclose $x^{B,d}$, since in the local expansion we want to approximate the potential inside $B$. Again both $y^{B,d}$ and $x^{B,d}$ should locate between $B$ and $\mathcal{F}(B)$. Two surfaces are shown in Figure \ref{fmulloc} as an example, with evaluation procedure.

The downward equivalent density satisfies:
\begin{equation}
\sum_{k \in Id(y^{B,d})} G(x_j^{B,d},y_k^{B,d}) \phi_k^{B,d}= q^{B,d}(x_j^{B,d}) = \sum_{i \in I^{\mathcal{F}(B)}_s} G(x_j^{B,d},y_i) I(y_i),\ \forall j \in Id(x^{B,d}),
\label{s2l}
\end{equation}
where $I_s^{\mathcal{F}(B)}$ is the index set of the source points in $\mathcal{F}(B)$, $Id(y^{B,d})$ is the index set of source points on $y^{B,d}$ and $Id(x^{B,d})$ is the index set of check points on $x^{B,d}$. Again a prescribed finite number (related to $m_0$) of index is chosen in $Id(\cdot)$.

\paragraph{M2M translation} M2M translation translates the upward equivalent density $\phi^{B,u}$ of a box, to the upward equivalent density $\phi^{\mathcal{P}(B),u}$ of its parent box $\mathcal{P}(B)$. The idea is similar to (\ref{s2m}), with an upward check surface of $\mathcal{P}(B)$ as $x^{\mathcal{P}(B),u}$, and the corresponding upward check potential $q^{\mathcal{P}(B),u}$. The equality is given as
\begin{equation}
\begin{split}
q^{\mathcal{P}(B),u}(x_j^{\mathcal{P}(B),u}) & =\sum_{k \in Id(y^{\mathcal{P}(B),u})} G(x_j^{\mathcal{P}(B),u},y_k^{\mathcal{P}(B),u}) \phi_k^{\mathcal{P}(B),u} \\
& =\sum_{i \in Id(y^{B,u})} G(x_j^{\mathcal{P}(B),u},y_i^{B,u}) \phi_i^{B,u},\qquad \forall j \in Id(x^{\mathcal{P}(B),u}).
\end{split}
\label{m2m}
\end{equation}

\begin{figure}[htb!]
\centering
\begin{tikzpicture}
\draw[line width = 1pt] (-0.8,-0.8) rectangle (0.8,0.8);
\draw[line width = 1pt] (-0.8,0) rectangle (0,0.8);
\node[circle, inner sep =0pt] at (-0.3,0.3) {$B$};
\node[circle, inner sep =0pt] at (0.3,-0.5) {$\mathcal{P}(B)$};
\draw[dashed, line width=1pt] (-0.96,-0.96) rectangle (0.96,0.96);
\draw[dashed, line width=1pt] (-0.88,-0.08) rectangle (0.08,0.88);
\draw[densely dotted, line width = 1pt] (-2.08,-2.08) rectangle (2.08,2.08);
\draw [red, line width=3.2pt,line cap=round,dash pattern=on 0pt off 0.32cm] (-0.88,-0.08) rectangle (0.08,0.88);
\draw [red, line width=4pt,line cap=round,dash pattern=on 0pt off 0.64cm] (-0.96,-0.96) rectangle (0.96,0.96);
\draw [blue, line width=4pt,line cap=round,dash pattern=on 0pt off 1.38667cm] (-2.08,-2.08) rectangle (2.08,2.08);
\node[circle, inner sep =0pt, text=blue] at (-1.4,1.7) {$x^{\mathcal{P}(B),u}$};
\node[circle, inner sep =0pt, text=red] at (-1,-1.3) {$y^{\mathcal{P}(B),u}$};
\node[circle, inner sep =0pt, text=red] at (0.45,0.4) {{\footnotesize $y^{B,u}$}};
\draw[->, line width = 1pt, blue] (0.08,0.1) -- (2.08,0.1) node[midway, above right] {{\footnotesize $(1)$}};
\draw[->, line width = 1pt, red] (0,-2.08) -- (0,-0.96) node[midway, right] {{\footnotesize $(2)$}};
%-----------------------------------------------------
\draw[line width = 1pt] (3.4,-1.6) rectangle (6.6,1.6);
\draw[line width = 1pt] (5, -1.6) -- (5, 1.6);
\draw[line width = 1pt] (3.4,0) -- (6.6, 0);
\draw[line width = 1pt] (3.4,0.8) -- (5, 0.8);
\draw[line width = 1pt] (4.2,0) -- (4.2, 1.6);
\draw[line width = 1pt] (5,-0.8) -- (6.6, -0.8);
\draw[line width = 1pt] (5.8,-1.6) -- (5.8, 0);
\node[circle, inner sep =0pt] at (3.8,1.2) {$A$};
\node[circle, inner sep =0pt] at (5.4,-0.4) {$B$};
\draw[dashed, line width=1pt] (3.32,0.72) rectangle (4.28,1.68);
\draw[densely dotted, line width=1pt] (4.92,-0.88) rectangle (5.88,0.08);
\draw[dashed, line width=1pt] (4.36,-1.44) rectangle (6.44,0.64);
\draw [red, line width=3.2pt,line cap=round,dash pattern=on 0pt off 0.32cm] (3.32,0.72) rectangle (4.28,1.68);
\draw [blue, line width=3.2pt,line cap=round,dash pattern=on 0pt off 0.32cm] (4.92,-0.88) rectangle (5.88,0.08);
\draw [red, line width=3.2pt,line cap=round,dash pattern=on 0pt off 0.693334cm] (4.36,-1.44) rectangle (6.44,0.64);
\draw[->, line width = 1pt, blue] (3.8,0.72) -- (4.92,-0.4) node[midway, left] {{\footnotesize $~(1)$}};
\draw[->, line width = 1pt, red] (5.88,-0.4) -- (6.44,-0.4) node[midway, below] {{\footnotesize $(2)$}};
\node[circle, inner sep =0pt, text=blue] at (5.4,-1.1) {$x^{B,d}$};
\node[circle, inner sep =0pt, text=red] at (2.9,0.6) {$y^{A,u}$};
\node[circle, inner sep =0pt, text=red] at (3.9,-1.3) {{$y^{B,d}$}};
%-------------------------------------------------
\draw[line width = 1pt] (9.2,-0.8) rectangle (10.8,0.8);
\draw[line width = 1pt] (9.2,0) rectangle (10,0.8);
\node[circle, inner sep =0pt] at (9.6,0.4) {$B$};
\node[circle, inner sep =0pt] at (10.2,-0.4) {$\mathcal{P}(B)$};
\draw[dashed, line width=1pt] (8.56,-0.64) rectangle (10.64,1.44);
\draw[dotted, line width=1pt] (9.12,-0.08) rectangle (10.08,0.88);
\draw[dashed, line width = 1pt] (7.92,-2.08) rectangle (12.08,2.08);
\draw [blue, line width=3.2pt,line cap=round,dash pattern=on 0pt off 0.32cm] (9.12,-0.08) rectangle (10.08,0.88);
\draw [red, line width=4pt,line cap=round,dash pattern=on 0pt off 0.693334cm] (8.56,-0.64) rectangle (10.64,1.44);
\draw [red, line width=4pt,line cap=round,dash pattern=on 0pt off 1.38667cm] (7.92,-2.08) rectangle (12.08,2.08);
\node[circle, inner sep =0pt, text=red] at (8.6,-1.8){$y^{\mathcal{P}(B),d}$};
\node[circle, inner sep =0pt, text=red] at (8.7,-1) {$y^{B,d}$};
\node[circle, inner sep =0pt, text=blue] at (9.6,1.1) {{\footnotesize $x^{B,d}$}};
\draw[->, line width = 1pt, blue] (12.08,0.1) -- (10.08,0.1) node[midway, above right] {{\footnotesize $(1)$}};
\draw[->, line width = 1pt, red] (9.6,-0.08) -- (9.6,-0.64) node[midway, left] {{\scriptsize $(2)$}};
\end{tikzpicture}
\caption{\label{m2mm2ll2l}Cross section of M2M translation (left subfigure), M2L translation (middle subfigure) and L2L translation (right subfigure). Dashed lines with red dots denote equivalent surfaces. Dotted lines with blue dots denote check surface. Step $(1)$ in blue arrow is the evaluation of potential on check surface, and step $(2)$ in red arrow is the calculation of equivalent density on equivalent surface. $\mathcal{P}(B)$ denotes parent box of $B$.}
\end{figure}
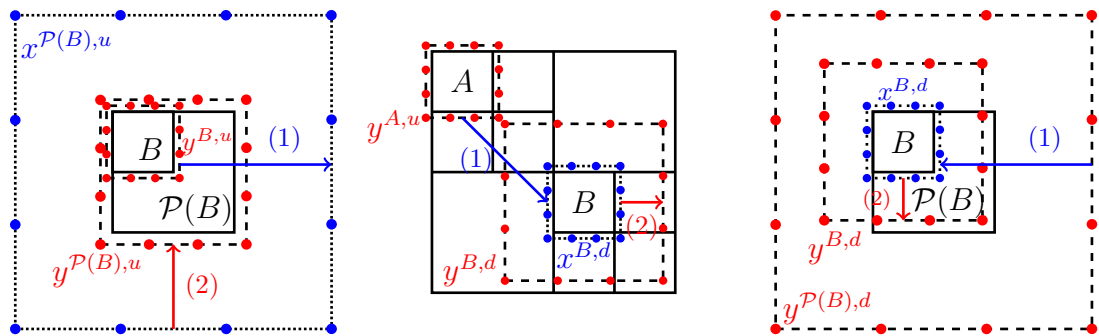

In the implementation, we first add the potential from the upward equivalent density of all children boxes to the check surface of the parent box, which is marked as blue arrow in the left-most subfigure of Figure \ref{m2mm2ll2l}. After accumulation from all children boxes to $q^{\mathcal{P}(B),u}$, we evaluate the upward equivalent density $\phi^{\mathcal{P}(B),u}$ which is marked as red arrow in the same subfigure. This implementation, which is adding potential to the check surface and then calculating the equivalent density from check potential, is also applied to the calculation of downward equivalent density. Therefore, we also indicate the implementation by blue and red arrows in other subfigures related to M2L and L2L translations in Figure \ref{m2mm2ll2l}.

\paragraph{M2L translation} Two boxes $A$ and $B$ are well-separated if $A \subset \mathcal{F}(B)$ and $B \subset \mathcal{F}(A)$. If two boxes $A$ and $B$ are in same size and well-separated, M2L translation can be used to translate the multipole expansion of $A$ to local expansion of $B$. In other words, M2L translation calculates the downward equivalent density of $B$ from the upward equivalent density of $A$, which accumulates the potential in $B$ from the source in $A$. See this procedure in Figure \ref{m2mm2ll2l}, which satisfies
\begin{equation}
\sum_{k \in Id(y^{B,d})} G(x_j^{B,d},y_k^{B,d}) \phi_k^{B,d}= q^{B,d}(x_j^{B,d}) = \sum_{i \in Id(y^{A,u})} G(x_j^{B,d},y_i^{A,u}) \phi_i^{A,u},\ \forall j \in Id(x^{B,d}).
\label{m2l}
\end{equation}
Since these two boxes $A$ and $B$ are in same size, fast Fourier transform can be used to speed up the calculation in (\ref{m2l}), as indicated in \cite{ying2004}.

\paragraph{L2L translation} For a box $B$, $\mathcal{F}(\mathcal{P}(B)) \subset \mathcal{F}(B)$, therefore L2L translation is used to calculation the local expansion of $B$ from the local expansion of $\mathcal{P}(B)$. This specifies the potential in $B$ from the source in $\mathcal{F}(\mathcal{P}(B))$. Similar as (\ref{m2m}), the equation is
\begin{equation}
\begin{split}
q^{B,d}(x_j^{B,d}) & =\sum_{k \in Id(y^{\mathcal{P}(B),d})} G(x_j^{B,d},y_k^{\mathcal{P}(B),d}) \phi_k^{\mathcal{P}(B),d} \\
& =\sum_{i \in Id(y^{B,d})} G(x_j^{B,d},y_i^{B,d}) \phi_i^{B,d},\qquad \forall j \in Id(x^{B,d}).
\end{split}
\label{l2l}
\end{equation}

\paragraph{Outline of the algorithm} The outline steps of the kernel-independent FMM is presented as follows:
\begin{enumerate}
\item[1.] Tree construction: to construct a hierarchical tree in pre-order traversal, such that each leaf box contains no more than $m$ source points.
\item[2.] Upward pass: to calculate the multipole expansion for leaf boxes, and use M2M translation for multipole expansion of all non-leaf boxes in a post-order traversal of the hierarchical tree.
\item[3.] Downward pass: for non-root boxes, use local expansion, M2L and L2L translations to accumulate the potential from far range in a pre-order traversal of the tree.
\item[4.] Target potential: for each leaf box in pre-order traversal of the tree, sum up the near interactions with the potential calculated in the last step, get the target potential.
\end{enumerate}

We would like to remark that we tested the backward stable pseudo-inverse trick indicated in \cite{malhotra2015pvfmm} for (\ref{integral}), and find few difference with results given by the original pseudo-inverse in \cite{ying2004} for our problems. Therefore, we implement the kernel-independent FMM by the original pseudo-inverse with prescribed number $m_0=6$ (number of equivalent source or check points at each side of the enclosing box) in this paper. We find that $m_0=6$ gives a good balance between accuracy and computational cost. Readers may consider increasing $m_0$ to obtain a more accurate result, or decreasing $m_0$ for a faster computation.

It should be emphasized that in order to capture the multiscale structure of streamers, a non-uniform mesh may be adopted in the simulations like in \cite{marskar20203d,bessieres2007}. The aforementioned fast multipole framework {\color{black}\cite{ying2004}} still works for non-uniform and unstructured meshes.

\section{Results and comparison for computing photoionization}\label{comparison}

In this section, we compare the performance, in terms of accuracy and efficiency, of different methods for the
evaluation of the photoionization $S_{\rm ph}$ defined in \eqref{integral} with \eqref{Ifun}. We take $V=V'=[0, x_d] \times [0, y_d] \times [0,z_d]$\, cm$^3$ and denote its center as $\vec{x}_0=(x_0,y_0,z_0)^T= (x_d/2, y_d/2, z_d/2)^T$\,cm. The box $V$ is partitioned uniformly by $n_x \times n_y \times n_z$ cells, with $n_x$, $n_y$ and $n_z$ the number of cells along $x$, $y$, $z$ directions, respectively.

\Table{\label{diffnames}Notations of several methods introduced in this paper.}
\br
\ns
Notation of method & Brief description \\
\mr
Classical Int & Direct calculation on (\ref{integral}), with (\ref{Ifun}) and (\ref{gfun}). \\
Helmholtz zero BC & Three-term summation on (\ref{sphHel}), \\
& by solving (\ref{Hel}) with zero boundary condition. \\
Helmholtz Int BC & Three-term summation on (\ref{sphHel}), \\
& by solving (\ref{Hel}) with integral boundary condition from (\ref{integral}). \\
SP$_3$ Larsen BC & Three-group summation on (\ref{rteSph}), \\
& by solving (\ref{phi1})--(\ref{phi2}) with boundary conditions (\ref{boundary1})--(\ref{boundary2}). \\
SP$_3$ Int BC & Three-group summation on (\ref{rteSph}), \\
& by solving (\ref{phi1})--(\ref{phi2}) with integral boundary condition (\ref{integral}). \\
FMM classical Int & Fast multipole method based on (\ref{integral}), with (\ref{Ifun}) and (\ref{gfun}). \\
\br
\end{tabular}
\end{indented}
\end{table}

For simplicity, different numerical methods to be compared
are summarized in Table \ref{diffnames}.
The numerical simulations were performed on the Tianhe2-JK cluster located at Beijing Computational Science Research Center. More details can be found at \verb|https://www.csrc.ac.cn/en/facility/cmpt/2015-05-07/8.html|. 
In our computations via the MPI parallelism, excepted stated otherwise, we always use 4 nodes with 20 cores in each node in the simulation.

The accuracy of different numerical methods is quantified by the following relative errors:
\begin{equation}
\begin{split}
&\mathcal{E}_V := \frac{ \| S_{\rm ph}^{\rm num}(\vec{x})-S_{\rm ph}^{\rm ref}(\vec{x}) \|_2}{\|S_{\rm ph}^{\rm ref}(\vec{x})\|_2}\times 100\%, \\
&\mathcal{E}_{\delta}(\vec{x}_0):= \frac{1}{N_{\text{tot}}} \sum_{|\vec{x} - \vec{x}_0 | \leq \delta} \frac{| S_{\rm ph}^{\rm num}(\vec{x})-S_{\rm ph}^{\rm ref}(\vec{x})|}{S_{\rm ph}^{\rm ref}(\vec{x})}\times 100\%,\\
\end{split}
\label{relativefun}
\end{equation}
where $\|\cdot\|_2$ is the standard discrete $L^2$-norm on $V$, $\vec{x}_0\in V$, $\delta>0$ is a constant to be fixed later, $S_{\rm ph}^{\rm ref}(\vec{x})$ is the reference result calculated by the (discrete) Classical Int method, $S_{\rm ph}^{\rm num}(\vec{x})$ is the numerical approximation by a numerical method, and $N_{\text{tot}}$ is the number of grid points located within a $\delta$-radius of $\vec{x}_0$. In fact, here $\mathcal{E}_V$ and $\mathcal{E}_{\delta}(\vec{x}_0)$ can be regarded as the global relative error over the whole domain $V$ and the local
relative error over a ball centered at $\vec{x}_0$ with a radius of $\delta$,
respectively.

\addi{The elliptic equations in Helmholtz or SP$_3$ methods are solved by the efficient multigrid-preconditioned FGMRES solver \cite{lin2018}, whose performance was shown in \cite{lin2018} for solving elliptic equations with either constant or varied coefficients.} 

\subsection{Gaussian emission source term with different sizes of the domain}\label{DiffR}

The first example is to compute the photoionization rate $S_{\rm ph}(\vec{x})$ in \eqref{integral} generated from a single Gaussian emission source, which is taken from \cite{bourdon2007}. The Gaussian ionization source $S_{i}(\vec{x})$ in \eqref{Ifun} is given as
\begin{equation}
S_{i}(\vec{x}) = 1.53 \times 10^{25} \exp\left(- \left( (x-x_0)^2 + (y-y_0)^2 + (z-z_0)^2\right) / \sigma^2 \right) \text{cm}^{-3}\,\text{s}^{-1},
\label{onegaussiansrc}
\end{equation}
where $\sigma>0$ is a constant to be fixed later.
The other physics parameters in \eqref{integral}-\eqref{gfun} are chosen as \cite{bourdon2007,segur2006}: $p_q =30$\,Torr, $p=760$\,Torr, $\xi=0.1$, $\omega/ \alpha = 0.6$, $p_{_{O_2}}=150$\,Torr. We take $\delta=5\sigma$
in \eqref{relativefun}.

We take a relatively small grid size $n_x=n_y=320$ and $n_z=160$ because direct computation of the classical integral (\ref{integral}) is too time-consuming even if parallel computing is utilized.

\begin{figure}[htb!]
\centering
\includegraphics[width=0.45\textwidth]{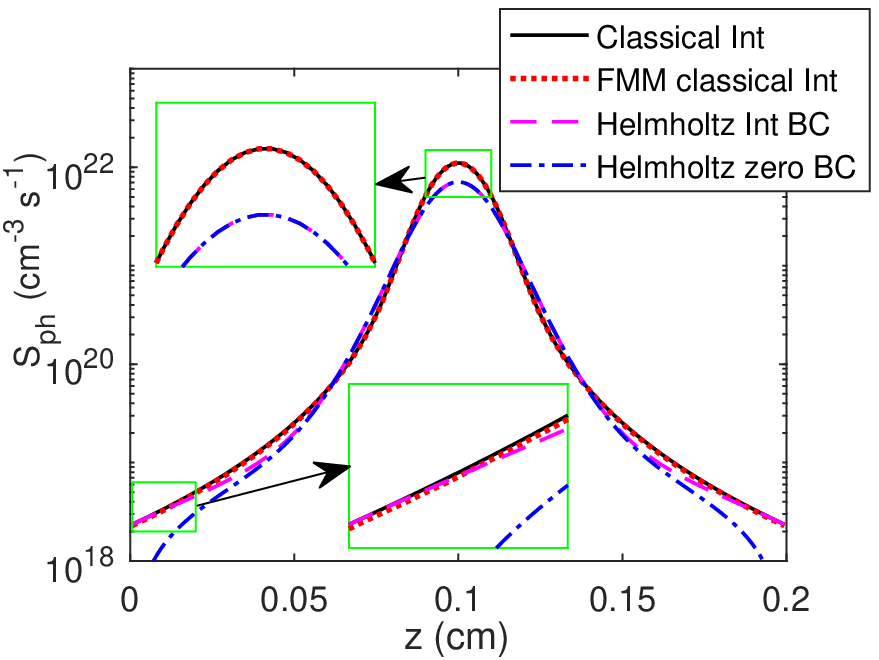}
\includegraphics[width=0.45\textwidth]{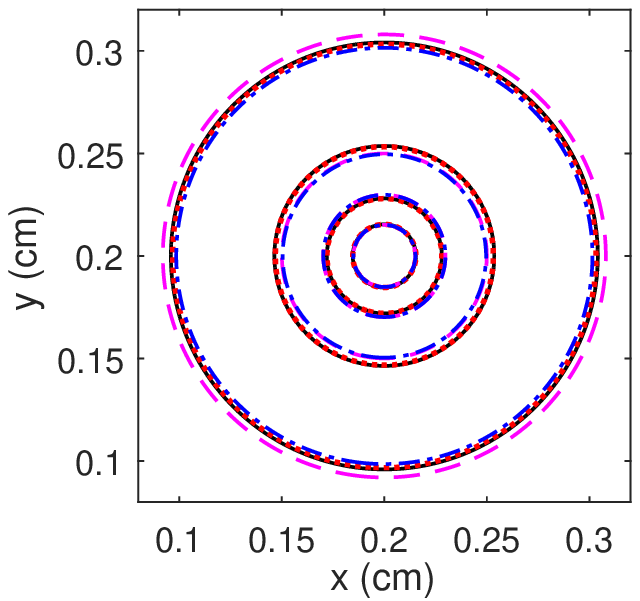}
\includegraphics[width=0.45\textwidth]{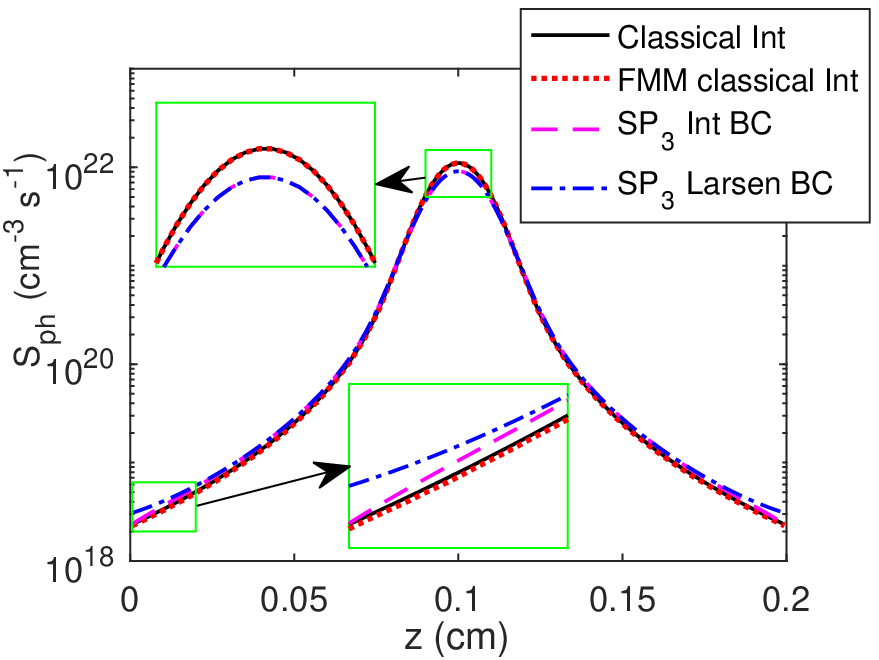}
\includegraphics[width=0.45\textwidth]{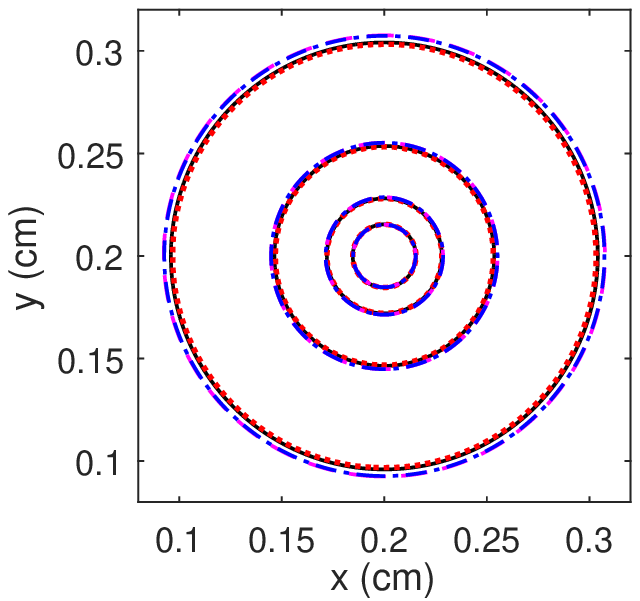}
\caption{\label{fonegauss001}Photoionization rate $S_{\rm ph}$ calculated from one Gaussian source in (\ref{onegaussiansrc}). $x_d=y_d=0.4$\,cm, $z_d=0.2$\,cm, $\sigma=0.01$\,cm. The figures in the left column are $S_{\rm ph}$ along line $x=y=0.2$\,cm, while the figures in the right column are contours of $S_{\rm ph}$ on the plane $z=0.1$\,cm, with the values of the contour lines being $2 \times 10^{18}$, $2 \times 10^{19}$, $2 \times 10^{20}$, $2 \times 10^{21}$\,cm$^{-3}$\,s$^{-1}$. The line color and format in the right-hand side subfigures are same as the one in the left-hand side in the same row.}
\end{figure}

\begin{figure}[htb!]
\centering
\includegraphics[width=0.45\textwidth]{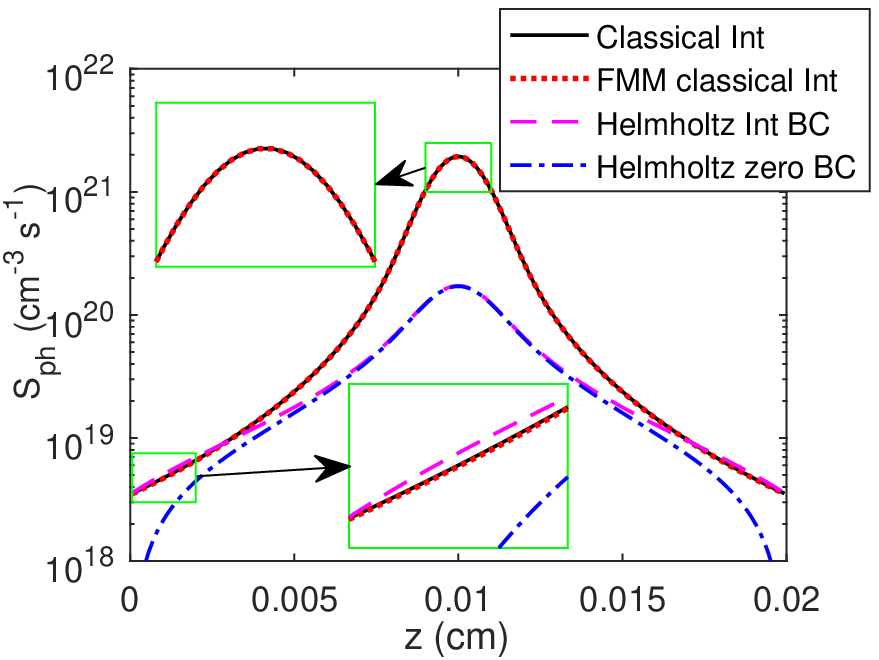}
\includegraphics[width=0.45\textwidth]{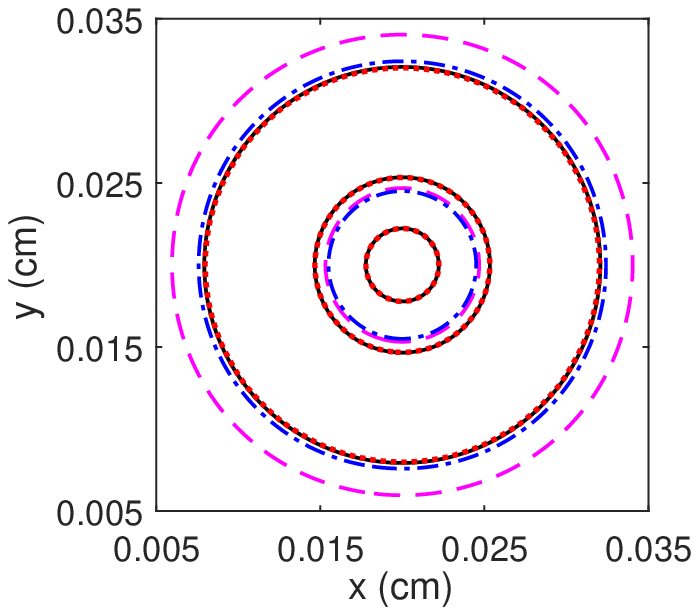}
\includegraphics[width=0.45\textwidth]{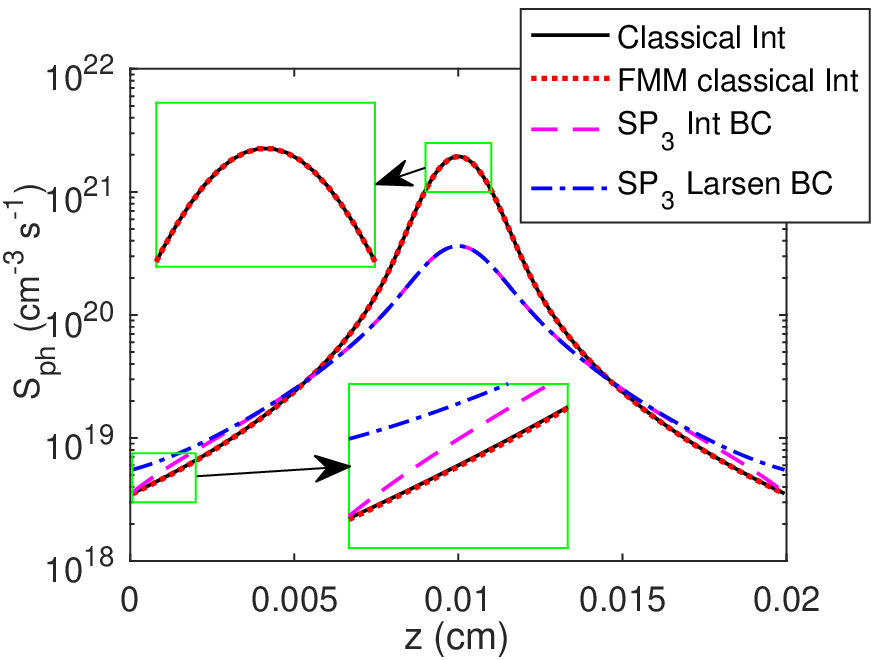}
\includegraphics[width=0.45\textwidth]{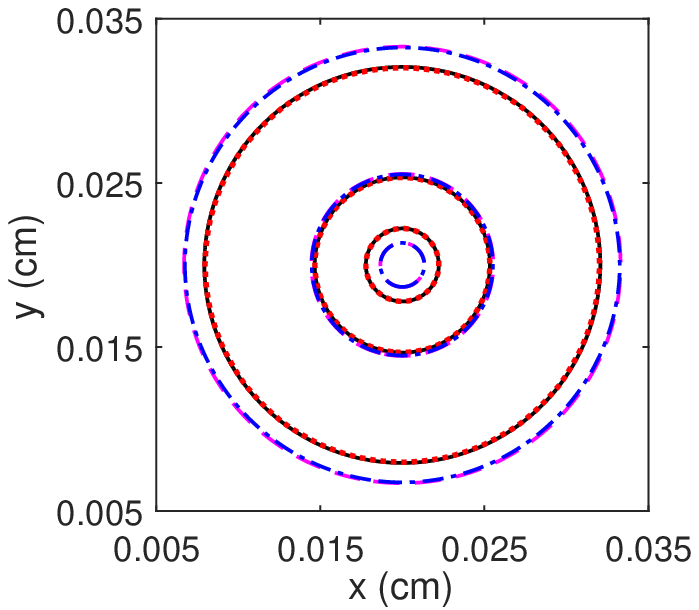}
\caption{\label{fonegauss0001}Photoionization rate $S_{\rm ph}$ calculated from one Gaussian source in (\ref{onegaussiansrc}). $x_d=y_d=0.04$\,cm, $z_d=0.02$\,cm, $\sigma=0.001$\,cm. Left-hand side subfigures are $S_{\rm ph}$ along line $x=y=0.02$\,cm, while right-hand side subfigures are contour line of $S_{\rm ph}$ on $z=0.01$\,cm plane, with contour values $2 \times 10^{18}$, $2 \times 10^{19}$, $2 \times 10^{20}$\,cm$^{-3}$\,s$^{-1}$. The line color and format in the right-hand side subfigures are same as the one in the left-hand side in the same row.}
\end{figure}

Similar to \cite{bourdon2007}, we demonstrate the influence of different ranges of $p_{_{O_2}}r$ by considering two different sizes of the domain $V$:
\begin{enumerate}
\item $x_d=y_d=0.4$\,cm, $z_d=0.2$\,cm, $\sigma=0.01$\,cm;
\item $x_d=y_d=0.04$\,cm, $z_d=0.02$\,cm, $\sigma=0.001$\,cm.
\end{enumerate}
The numerical results are shown in Figures \ref{fonegauss001} and \ref{fonegauss0001}, and the relative errors are then shown in Tables \ref{tonegaussiansrc001} and \ref{tonegaussiansrc0001}.

\Table{\label{tonegaussiansrc001}Time usage and relative error of methods indicated in Table \ref{diffnames}, for the case of single Gaussian source $x_d=y_d=0.4$\,cm, $z_d=0.2$\,cm, $\sigma=0.01$\,cm. \addi{$\vec{x}_0 = (0.2, 0.2, 0.1)^T$\,cm and $\delta = 5 \sigma$.}}
\br
Method & Time usage (s) & $\mathcal{E}_V$ & $\mathcal{E}_{\delta}(\vec{x}_0)$\\
\br
Classical Int & 184248 & --- & --- \\
FMM classical Int & 27.1897 & 0.21\% & 1.30\% \\
Helmholtz zero BC & 3.66044 & 25.33\% & 16.37\% \\
Helmholtz Int BC & 3.76133+4606.20$^{\rm a}$ & 25.32\% & 15.54\% \\
$SP_3$ Larsen BC & 12.9268 & 12.05\% & 8.49\% \\
SP$_3$ Int BC & 7.30268+4606.20$^{\rm a}$ & 12.05\% &8.53\% \\
\br
\end{tabular}
\item[] $^{\rm a}$time usage to compute the boundary values, which is estimated from Classical Int method, with multiplication to a factor $2(n_x \times n_y + n_x \times n_z + n_y \times n_z)/(n_x \times n_y \times n_z)$.
\end{indented}
\end{table}

As it is clearly shown in Figure \ref{fonegauss001} and Figure \ref{fonegauss0001}, the FMM classical Int method always gives the most accurate results, especially for the smaller domain. In all the figures, the lines for ``FMM classical Int'' almost coincide with the lines for ``Classical Int''. The deviations of the solutions of the other four methods from the reference results are clearly observable, especially in the central area where the peak locates. Near the boundaries, the methods based on modified Helmholtz equations and SP$_3$ equations are accurate only when the boundary values are computed from direct integration. Tables \ref{tonegaussiansrc001} and \ref{tonegaussiansrc0001} also show the superiority of the FMM method in terms of accuracy. Its relative error is one or two orders of magnitude less than other methods.

Regarding the efficiency, it should be noted that both Helmholtz Int BC method and SP$_3$ Int BC method take the boundary values from the Classical Int method, and the time to compute the boundary conditions is also included in Table \ref{tonegaussiansrc001} and Table \ref{tonegaussiansrc0001} for a fair comparison. Both tables show that the FMM classical Int method is significantly faster than the Classical Int method. In fact, the integration only for the boundary nodes is already much more expensive than the FMM classical Int method. For the three efficient methods, including FMM classical Int, Helmholtz zero BC, and SP$_3$ Larsen BC, their computational times have similar magnitudes, and the speed-accuracy trade-off can be observed, meaning that higher computational cost yields better accuracy. Nevertheless, the remarkably lower numerical error and the mildly higher computational cost of the FMM classical Int method indicate its outstanding competitiveness among all the approaches for computing the photoionization rates. Additionally, the time usages of FMM classical Int method are stable for different problem settings, while that of SP$_3$ Larsen BC method varies significantly (see Tables  \ref{tonegaussiansrc001} and \ref{tonegaussiansrc0001}).

\Table{\label{tonegaussiansrc0001}Time usage and relative error of methods indicated in Table \ref{diffnames}, for the case of one Gaussian source $x_d=y_d=0.04$\,cm, $z_d=0.02$\,cm, $\sigma=0.001$\,cm. \addi{$\vec{x}_0 = (0.02, 0.02, 0.01)^T$\,cm and $\delta = 5 \sigma$.}} 
\br
Method & Time usage (s) & $\mathcal{E}_V$ & $\mathcal{E}_{\delta}(\vec{x}_0)$\\
\br
Classical Int & 183761 & --- & --- \\
FMM classical Int & 27.6406 & 0.22\% & 0.53\% \\
Helmholtz zero BC & 3.87327 & 83.26\% & 48.03\% \\
Helmholtz Int BC & 4.09442+4594.03$^{\rm a}$ & 82.96\% & 44.82\% \\
SP$_3$ Larsen BC & 17.3571 & 68.92\% & 16.67\% \\
SP$_3$ Int BC & 7.88867+4594.03$^{\rm a}$ & 68.88\% & 16.53\% \\
\br
\end{tabular}
\item[] $^{\rm a}$ Estimated from the time usage of Classical Int method, with multiplication to a factor $2(n_x \times n_y + n_x \times n_z + n_y \times n_z)/(n_x \times n_y \times n_z)$.
\end{indented}
\end{table}

\subsection{Gaussian emission source with different pressures}\label{DiffP}

The second example is to compute the photoionization rate $S_{\rm ph}(\vec{x})$ in \eqref{integral} generated from a single Gaussian radiation source, 
which is taken from \cite{fvmrte2008},
in order to test the effect of the partial pressure of oxygen $p_{_{O_2}}$ 
in the kernel $g$ given in \eqref{gfun}. The Gaussian source of radiation
$I$ in \eqref{Ifun} is taken as \cite{fvmrte2008}
\begin{equation}
I(\vec{x}) = 4 \pi \xi c \exp \left[ - \frac{ (x-x_0)^2 + (y-y_0)^2+(z-z_0)^2}{\sigma^2} \right] \text{cm}^{-3}\,\text{s}^{-1},
\label{po2init}
\end{equation}
where $c$ is the speed of light. Similar to \cite{fvmrte2008}, we take $\xi=0.1$, $\sigma=0.01$\,cm, $c=2.99792458 \times 10^{10}$\,cm$\cdot$s$^{-1}$, $\delta=5\sigma$ and $V=[0,0.25] \times [0,0.25] \times [0,1.4]$\,cm$^{3}$. We fix the ratio of the partial pressure of oxygen and the air pressure $p_{_{O_2}}/p=0.2$ in this example. Finally, the box $V$ is uniformly partitioned by $256 \times 256 \times 320$ cells.

In this example, when the partial pressure of oxygen $p_{_{O_2}}$ in \eqref{gfun} is lower, the photoionization rate decays slower as the distance from the emission source increases. Therefore, we would like to test the performance of different methods under different pressures $p_{_{O_2}}$. The robustness of the methods under different pressures is of great significance in practical applications such as sprite discharges \cite{liu2015sprite,liu2004effects}. 

\begin{figure}[htb!]
\centering
\includegraphics[width=0.45\textwidth]{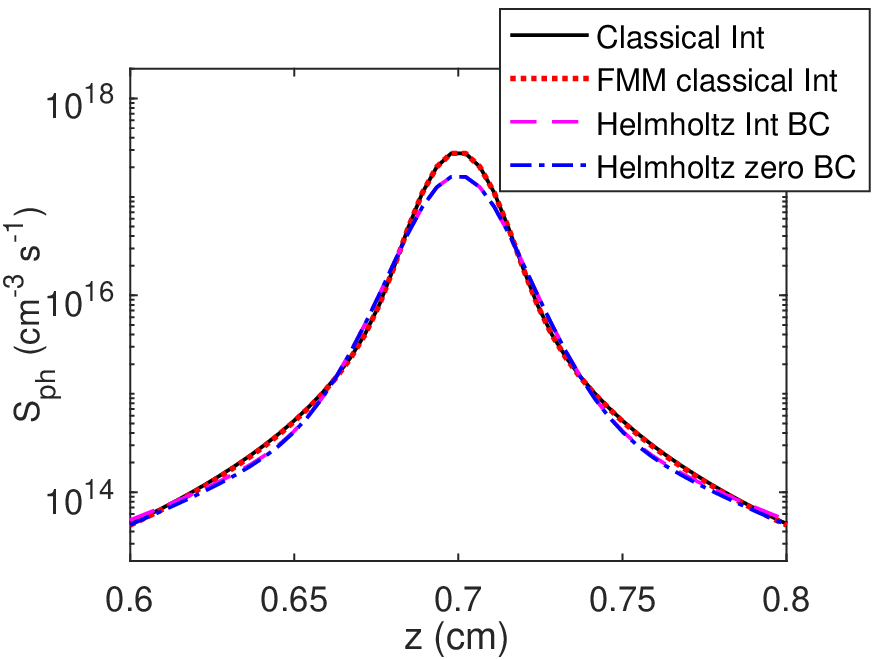}
\includegraphics[width=0.45\textwidth]{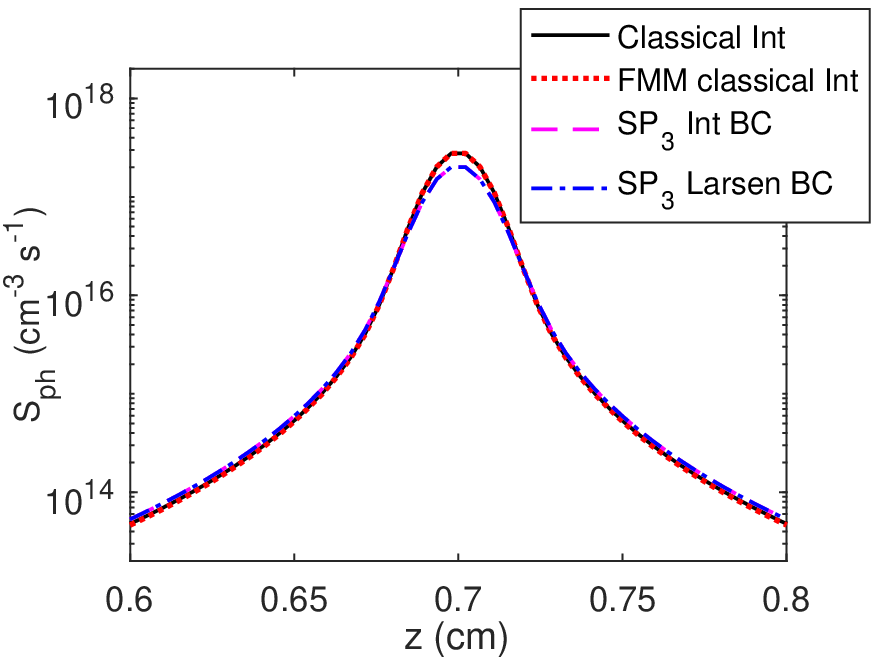}
\caption{\label{fdiffpi}Photoionization rate $S_{\rm ph}$ along line $x=y=0.125$\,cm, calculated from one Gaussian in (\ref{po2init}) with $p_{_{O_2}}=160$\,Torr.}
\end{figure}

\begin{figure}[htb!]
\centering
\includegraphics[width=0.45\textwidth]{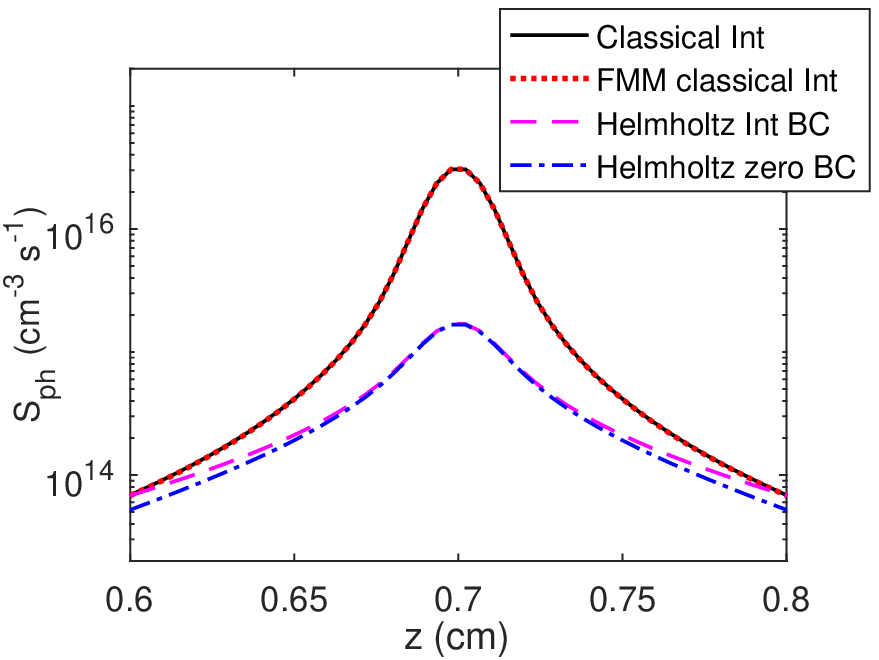}
\includegraphics[width=0.45\textwidth]{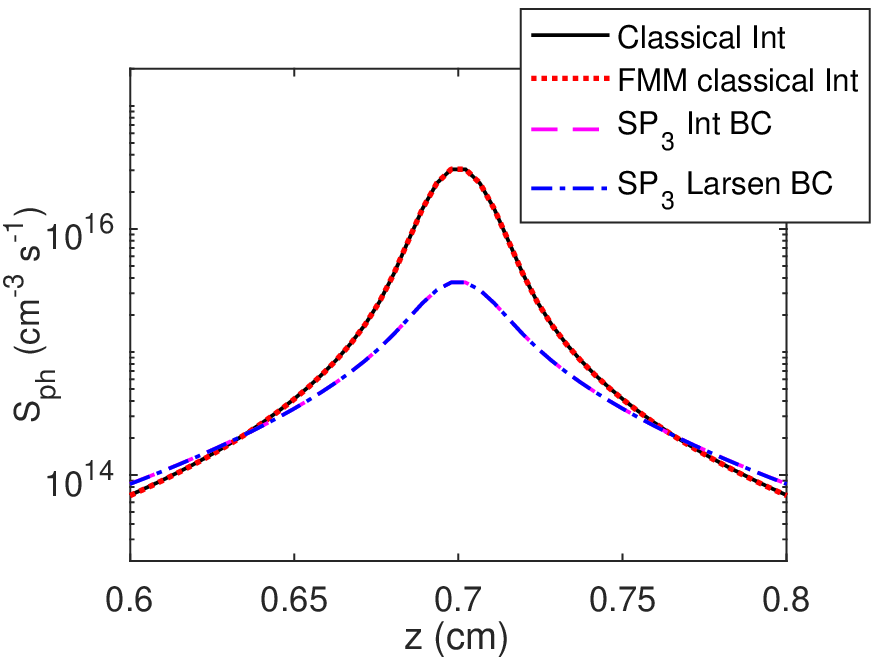}
\caption{\label{fdiffpiii}Photoionization rate $S_{\rm ph}$ along line $x=y=0.125$\,cm, calculated from one Gaussian in (\ref{po2init}) with $p_{_{O_2}}=10$\,Torr.}
\end{figure}

For comparison purpose, two different pressures are considered: (i) $p_{_{O_2}}=160$\,Torr; (ii) $p_{_{O_2}}=10$\,Torr. The photoionization rate along the central vertical line is plotted in Figures \ref{fdiffpi} and \ref{fdiffpiii}, and the time usage and the numerical error are shown in Tables \ref{tdiffpi} and \ref{tdiffpiii}.

\Table{\label{tdiffpi}Time usage and relative error of methods indicated in Table \ref{diffnames}, for the case of one Gaussian in (\ref{po2init}) with $p_{_{O_2}}=160$\,Torr. \addi{$\vec{x}_0$ is the center of domain $V$ and $\delta = 0.05$\,cm.}} 
\br
Method & Time usage (s) &$\mathcal{E}_V$ &$\mathcal{E}_{\delta}(\vec{x}_0)$ \\
\br
Classical Int & 292196 & --- & --- \\
FMM classical Int & 24.7371 & 0.15\% & 1.24\% \\
Helmholtz zero BC & 5.59405 & 31.93\% & 16.49\% \\
Helmholtz Int BC & 5.60994+6391.79$^{\rm a}$ & 31.93\% & 15.95\% \\
$SP_3$ Larsen BC & 17.6286 & 21.31\% & 8.74\% \\
SP$_3$ Int BC & 11.2337+6391.79$^{\rm a}$ & 21.31\% & 8.73\% \\
\br
\end{tabular}
\item[] $^{\rm a}$ Estimated from the time usage of Classical Int method, with multiplication to a factor $2(n_x \times n_y + n_x \times n_z + n_y \times n_z)/(n_x \times n_y \times n_z)$.
\end{indented}
\end{table}

\Table{\label{tdiffpiii}Time usage and relative error of methods indicated in Table \ref{diffnames}, for the case of one Gaussian in (\ref{po2init}) with $p_{_{O_2}}=10$\,Torr. \addi{$\vec{x}_0$ is the center of domain $V$ and $\delta = 0.05$\,cm.}} 
\br
Method & Time usage (s) &$\mathcal{E}_V$ &$\mathcal{E}_{\delta}(\vec{x}_0)$\\
\br
Classical Int & 292426 & --- & --- \\
FMM classical Int & 24.8619 & 0.19\% & 0.44\% \\
Helmholtz zero BC & 6.17337 & 88.73\% & 65.87\% \\
Helmholtz Int BC & 6.01811+6396.82$^{\rm a}$ & 88.29\% & 63.03\% \\
SP$_3$ Larsen BC & 23.5598 & 77.74\% & 35.11\% \\
SP$_3$ Int BC & 12.2226+6396.82$^{\rm a}$ & 77.71\% & 35.16\% \\
\br
\end{tabular}
\item[] $^{\rm a}$ Estimated from the time usage of Classical Int method, with multiplication to a factor $2(n_x \times n_y + n_x \times n_z + n_y \times n_z)/(n_x \times n_y \times n_z)$.
\end{indented}
\end{table}

Figure \ref{fdiffpi} shows that in the high-pressure case, all methods give similar results despite obvious mismatch of the peak values. As the pressure decreases, the discrepancy between different methods becomes more obvious, as shown in Figure \ref{fdiffpiii}. The curves given by the Helmholtz and SP$_3$ methods (with both boundary conditions) deviate significantly from the curves of Classical Int method in the low-pressure case. On the contrary, the results of the FMM classical Int method and the reference Classical Int method are in good agreement regardless of the air pressure. The values of the errors provided in Tables \ref{tdiffpi} and \ref{tdiffpiii} again show the advantage of the FMM classical Int method in accuracy. In fact, for the case of low air pressure, the time used by the FMM classical Int is quite close to the method of SP$_3$ Larsen BC.

\begin{figure}[htb!]
\centering
\includegraphics[width=0.48\textwidth]{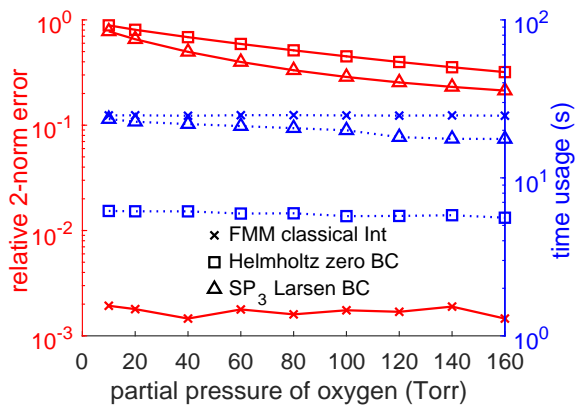}
\caption{\label{pressureCom}Error $\mathcal{E}_V$ (red color, solid line) and time usages (blue color, dotted line) of FMM classical Int, Helmholtz zero BC and SP$_3$ Larsen BC methods with different air pressures.}
\end{figure}

In order to see the relationship between the error and computational time with respect to different pressures, we compute more numerical examples under the same settings with different partial pressures of oxygen ranging from $10$\,Torr to $160$\,Torr. Results for the three most efficient method, i.e., FMM classical Int, Helmholtz zero BC and SP$_3$ Larsen BC methods, are plotted in Figure \ref{pressureCom}. The FMM classical Int method always provides the most accurate results for all pressures, and the error is basically stable as the pressure varies. For the other two methods, the error increases as pressure decreases. Moreover, the global relative error $\mathcal{E}_V$ 
of the FMM classical Int method are, in general, at least two order of magnitudes less than those of the other two methods. The time usage of the FMM classical Int method is also independent of pressure while the computation times of the other two methods increase slightly as the pressure becomes lower. As the pressure decreases, the time cost between the FMM classical Int method and the SP$_3$ Larsen BC method trends to be the same.

\addi{It should be mentioned here that the results of Helmholtz methods and SP$_3$ methods in this example are based on the fitting coefficients in Tables \ref{Helcoeff} and \ref{threegroupcoeff}. These results may be better if the coefficients are fit according to different $p_{_{O_2}}$ in this example. However, the large approximating errors, which is two order of magnitudes larger compared to the FMM classical Int method, over all ranges of $p_{_{O_2}}$ in Figure \ref{pressureCom} imply the FMM classical Int method would give more accurate results even when better fitting is applied.}

\addi{
\subsection{Multi-peak emission source}\label{chaptermultipeak}
The examples in previous two subsections compute the photoionization from one single Gaussian emission source, which is typical for comparison of photoionization and was used in \cite{bourdon2007, fvmrte2008}. In order to see the performance of different methods on other shapes of emission sources, we designed the third example, which computes the photoionization rate $S_{\rm ph}$ generated from a multi-peak emission source. 

All the parameters including physics parameters, simulation domain and grid size are taken identical to the example (i) in Section \ref{DiffR} except changing the source term $S_{i}(\vec{x})$ from \eqref{onegaussiansrc} to
\begin{equation}
S_i(\vec{x}) = \left\{
\begin{aligned}
& 1.53 \times 10^{25} \sin^2(50 \pi x) \sin^2(50 \pi y) \sin^2(50 \pi z) \text{cm}^{-3}\,\text{s}^{-1},\\
& ~~~\hspace{2cm} \vec{x} \in [0.25x_d, 0.75x_d] \times [0.25y_d, 0.75y_d] \times [0.25z_d, 0.75z_d],\\
& 0, \hspace{2cm} \vec{x} \notin [0.25x_d, 0.75x_d] \times [0.25y_d, 0.75y_d] \times [0.25z_d, 0.75z_d],
\end{aligned}
\right.
\label{multipeaksource}
\end{equation}
where $x_d=y_d=0.4\,$cm and $z_d=0.2$\,cm. The emission source in \eqref{multipeaksource} has hundreds of peaks, and has discontinuity inside the simulation domain $V$. 

The numerical results are depicted in Figure \ref{fmultipeak}, and the relative errors as well as the time usage are given in Table \ref{tmultipeak}.

\begin{figure}[htb!]
\centering
\includegraphics[width=0.45\textwidth]{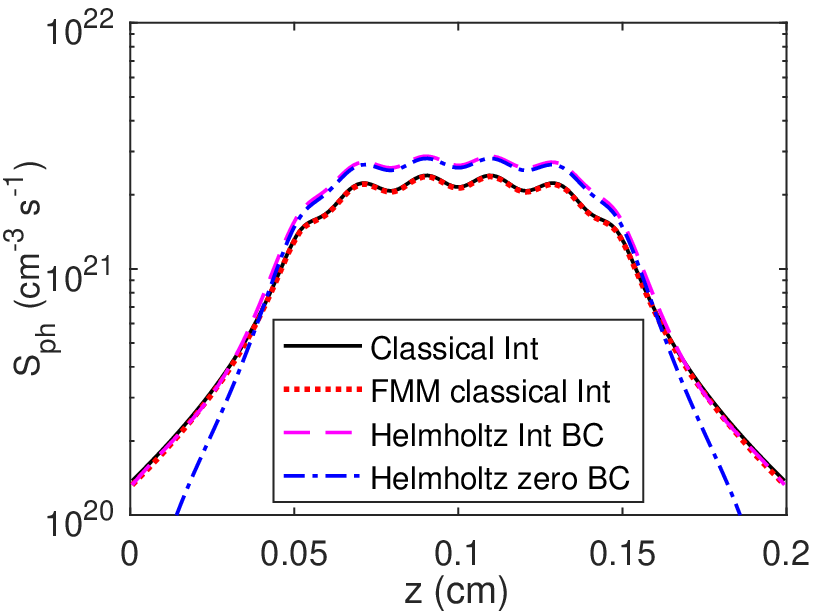}
\includegraphics[width=0.45\textwidth]{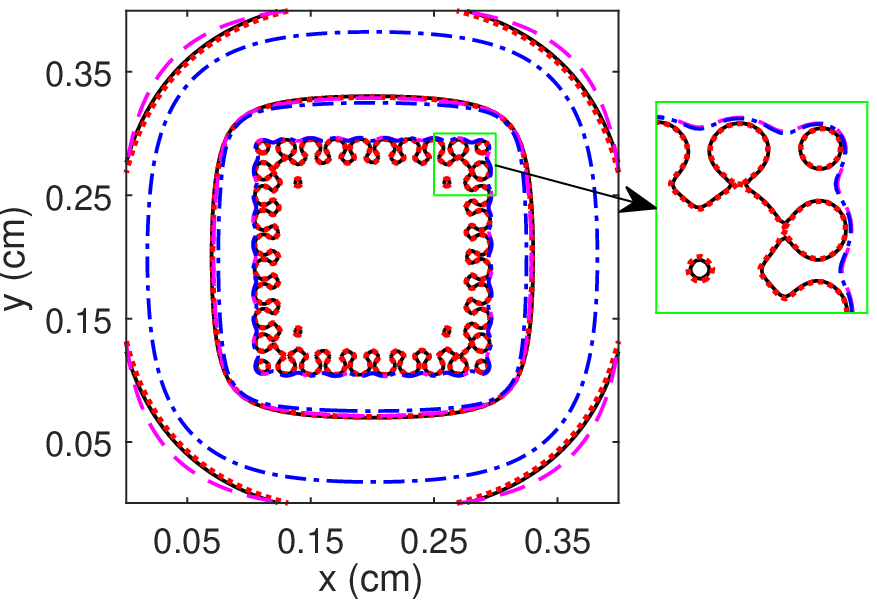}
\includegraphics[width=0.45\textwidth]{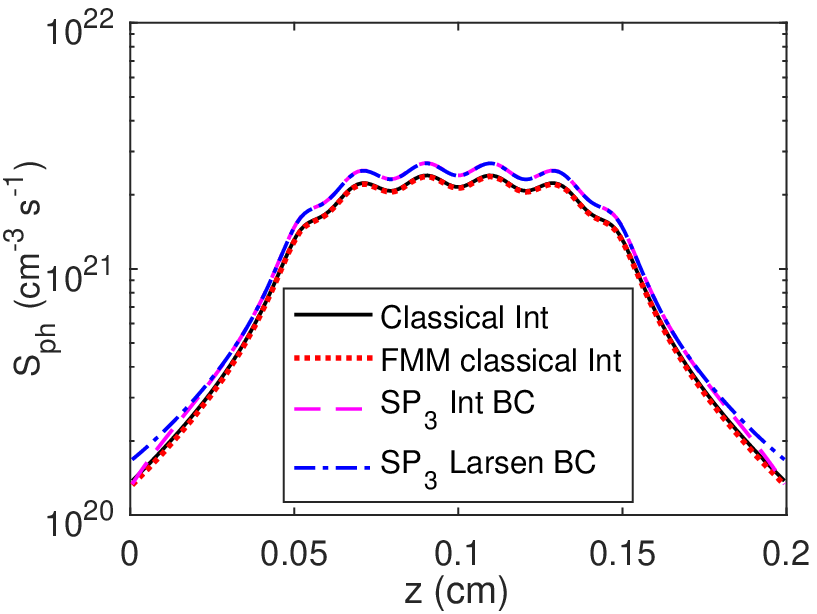}
\includegraphics[width=0.45\textwidth]{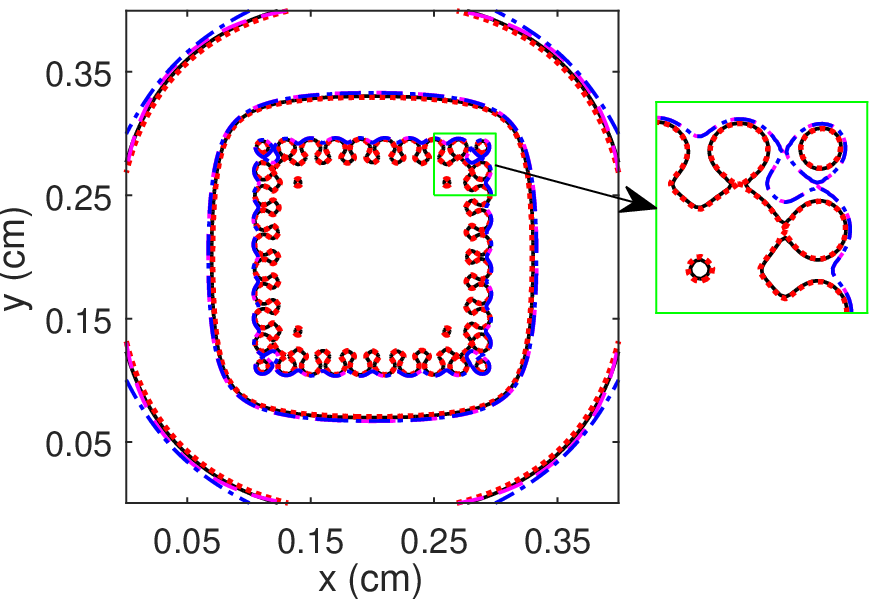}
\caption{\addi{\label{fmultipeak}Photoionization rate $S_{\rm ph}$ calculated from multi-peak source in \eqref{multipeaksource}. $x_d=y_d=0.4$\,cm, $z_d=0.2$\,cm. The figures in the left column are $S_{\rm ph}$ along line $x=y=0.2$\,cm, while the figures in the right column are contours of $S_{\rm ph}$ on the plane $z=0.1$\,cm, with the values of the contour lines being $2 \times 10^{19}$, $2 \times 10^{20}$, $2 \times 10^{21}$\,cm$^{-3}$\,s$^{-1}$. The line color and format in the right-hand side subfigures are same as the one in the left-hand side in the same row.}}
\end{figure}

\Table{\addi{\label{tmultipeak}Time usage and relative error of methods indicated in Table \ref{diffnames}, for the case of multi-peak source in Section \ref{chaptermultipeak}. $\vec{x}_0 = (0.2, 0.2, 0.1)^T$\,cm and $\delta = 0.05$\,cm.}}
\br
\addi{Method} & \addi{Time usage (s)} & \addi{$\mathcal{E}_V$} & \addi{$\mathcal{E}_{\delta}(\vec{x}_0)$} \\
\br
\addi{Classical Int} & \addi{171714} & --- & --- \\
\addi{FMM classical Int} & \addi{17.1896} & \addi{0.69\%} & \addi{0.82\%} \\
\addi{Helmholtz zero BC} & \addi{3.22809} & \addi{26.13\%} & \addi{14.13\%} \\
\addi{Helmholtz Int BC} & \addi{3.51407+4292.85$^{\rm a}$} & \addi{25.67\%} & \addi{15.06\%} \\
\addi{SP$_3$ Larsen BC} & \addi{14.5592} & \addi{13.84\%} & \addi{8.87\%} \\
\addi{SP$_3$ Int BC} & \addi{6.91124+4292.85$^{\rm a}$} & \addi{13.82\%} & \addi{8.98\%} \\
\br
\end{tabular}
\addi{\item[] $^{\rm a}$time usage to compute the boundary values, which is estimated from Classical Int method, with multiplication to a factor $2(n_x \times n_y + n_x \times n_z + n_y \times n_z)/(n_x \times n_y \times n_z)$.}
\end{indented}
\end{table}

Figure \ref{fmultipeak} illustrates the FMM classical Int method still gives the most accurate results when the source contains hundreds of peaks. All the lines of the FMM classical Int method greatly coincide with the lines for the Classical Int method, especially near the center of domain where the photoionization rate $S_{\rm ph}$ has several peaks. This coincidence still holds when the contour line of $2 \times 10^{19}$\,cm$^{-3}$\,s$^{-1}$ are highly oscillatory, which can be observed on the right subfigures of Figure \ref{fmultipeak}. On the other hand, the deviation of the other four approximation methods are clearly observable near the central region, and the contours of these approximation methods could not well follow the oscillatory contour line of $2 \times 10^{19}$\,cm$^{-3}$\,s$^{-1}$ for the the Classical Int method. The accurate approximation of the FMM classical Int method can also be observed quantitatively in Table \ref{tmultipeak}, where the errors of FMM Int are at least one order of magnitude less than other approximation methods.

The results of efficiency in Table \ref{tmultipeak} are similar to the results in Table \ref{tonegaussiansrc001}. Compared with the single Gaussian emission source used in Table \ref{tonegaussiansrc001}, the time usage of the FMM classical Int method as well as its difference to the other two efficient methods (Helmholtz zero BC and SP$_3$ Larsen BC) becomes smaller in this multi-peak example, and the FMM classical Int method still gives remarkably lower numerical error which verifies again the robustness of the FMM classical Int method.
}
\section{Results and comparison for computing streamer discharges}\label{comparison1}

To further compare their performances of different methods for treating the 
photoionization $S_{\rm ph}(\vec{x})$ in \eqref{integral}, we study the dynamics of streamers with photoionization, where $S_{\rm ph}$ appears as the source term of the transport of charged particles. The governing equations for streamer discharges are given as \cite{lin2018,bessieres2007}:
\begin{align}
\left\{ \begin{aligned}
& \frac{\partial n_e}{\partial t} - \nabla \cdot (\mu_e \vec{E} n_e) - \nabla \cdot (D_e \nabla n_e) = S_i + S_{\rm ph}, \\
& \frac{\partial n_p}{\partial t} + \nabla \cdot (\mu_{p} \vec{E} n_p) = S_i + S_{\rm ph}, \\
& {- \Delta \phi} = \frac{e}{\varepsilon_0} (n_p-n_e), \qquad \vec{E} = -\nabla \phi,
\end{aligned}
\right.
\label{eM}
\end{align}
where $e$ and $\varepsilon_0$ are the elementary charge and the vacuum dielectric permittivity, respectively; $n_e:=n_e(\vec{x},t)$ and $n_p:=n_p(\vec{x},t)$ are the densities of electrons and positive ions, respectively; $\mu_e$ and $\mu_p$ are the mobility coefficients for electrons and positive ions, respectively; {\color{black}$D_e$ is the diffusion coefficient}; $\phi$ and $\vec{E}$ denote the electric potential and electric field, respectively. Here the photoionization rate $S_{\rm ph}$ is given in \eqref{integral}-\eqref{gfun} with $S_i$ defined as
\begin{equation}
S_i=\mu_e n_e \alpha|\vec{E}|, \qquad 
\hbox{with}\quad \alpha:=\alpha(|\vec{E}|)=5.7p \exp(-260p/|\vec{E}|)\,{\rm cm}^{-1},
\end{equation}
while $\alpha$ is taken from \cite{dhali1987two}, $p$ is the air pressure, $n_e$ and $\vec{E}$ are the solution of \eqref{eM}.

The streamer discharge between two parallel plates are used for comparision. The computational domain $V$ is set to be a three-dimensional axis-aligned hyper-rectangle. For the Poisson equation, the Dirichlet boundary conditions are applied on the two faces perpendicular to the $z$ axis, i.e., $\phi=\phi_0$ on the upper face and $\phi=0$ on the bottom face; and the homogeneous Neumann boundary conditions are applied on the other four faces. Homogeneous Neumann boundary conditions are assigned on all boundaries for $n_e$ and the inflow boundaries for $n_p$. The initial value is set as 
\begin{equation}
\label{initst}
n_e(\vec{x},t=0)=n_p(\vec{x},t=0)=\tilde{n}_0(\vec{x}).
\end{equation} 
The parameteres are selected as follows \cite{lin2018,dhali1987two}: $\mu_e=2.9\times 10^5/p$ cm$^2/$(V$\cdot$s), $\mu_p=2.6 \times 10^3/p$ cm$^2/$(V$\cdot$s); $p=760$\,Torr, $\phi_0=52$\,kV. {\color{black} $D_e = 1800 $ cm$^2/$s \cite{De}}. Two constants $e$ and $\varepsilon_0$ are the elementary charge and permittivity of vacuum, respectively. The other physics parameters in \eqref{integral}-\eqref{gfun} are chosen as \cite{bourdon2007,segur2006}: $p_q =30$\,Torr, $\xi=0.1$, $\omega/ \alpha = 0.6$. \addii{It should be noted that for different percentages of oxygen in the mixture, the coefficients should be chosen accordingly. For convenience and comparison, we simply choose the fixed coefficients in Nitrogen, and the presented results in this section are numerical experiments which simply consider the major mechanism of the streamer. 
}

The numerical method for discretizing \eqref{eM} follows our previous work \cite{lin2018, tmag2020}. For spatial discretization, the second-order MUSCL method with Koren limiter is applied to the drift terms, and the central difference scheme is chosen for the diffusion term. The second-order explicit method is adopted for the time integration of (\ref{eM}) \cite{tmag2020}. The multigrid-preconditioned FGMRES is used as the efficient algebraic elliptic solver to solve the Poisson equation in (\ref{eM}) iteratively. The iteration terminates when the relative residual is less than $10^{-8}$. The other elliptic equations (\ref{Hel}), (\ref{phi1}) and (\ref{phi2}) are solved by the same algebraic elliptic solver, with a weaker stopping condition that the relative residual is less than $10^{-6}$.

\subsection{Double-headed streamers in air}
In this subsection, we consider the interaction of two double-headed streamers and compare the numerical results of the three most efficient methods: FMM classical Int method, Helmholtz zero BC method and SP$_3$ Larsen BC method. 

The initial value $\tilde{n}_0$ in \eqref{initst} is taken as
\begin{align*}
\tilde{n}_0(\vec{x}) = 10^{14} \Big( & \exp\left(- \left( (x-0.22)^2 + (y-0.25)^2 + (z-0.41)^2\right) / (0.03)^2 \right) \\
+ & \exp\left(- \left( (x-0.28)^2 + (y-0.25)^2 + (z-0.59)^2\right) / (0.03)^2 \right) \Big) \text{cm}^{-3}.
\end{align*}
The computational domain is fixed as $V=[0, 0.5] \times [0, 0.5] \times [0,1]$\,cm$^3$, which is partitioned by a uniform grid of $512 \times 512 \times 1280$ cells. The time step is chosen as $\Delta t=2.5\times 10^{-3}$\,ns. In order to see the interaction with respect to different $p_{_{O_2}}$, we pick two values as \addi{$p_{_{O_2}}=0.1$\,Torr} and $p_{_{O_2}}=150$\,Torr, respectively, in our simulations.

\addi{ For a proper approximation, we fit new groups of coefficients for $10^{-4}<p_{_{O_2}}r<0.2$ for the simulation using Helmholtz zero BC method and SP$_3$ Larsen BC method with $p_{_{O_2}}=0.1$ \,Torr. The new coefficients for Helmholtz zero BC method replace Table \ref{gexp} by: $C_{1}=9.7496$, $\lambda_1=8.2035$, $C_{2}=56.065$, $\lambda_2=61.588$, $C_{3}=565.99$ and $\lambda_3=494.18$. The new coefficients for SP$_3$ Larsen BC method replace Table \ref{rtetoint} by: $A_{1}=0.019219$, $\lambda_1=0.000064638$, $A_{2}=0.10796$, $\lambda_2=0.10189$, $A_{3}=0.35854$ and $\lambda_3=1.3474$. These two groups of coefficients are only used for $p_{_{O_2}}=0.1$\,Torr in this paper.}

\begin{figure}[htb!]
	\centering
	\includegraphics[width=0.48\textwidth]{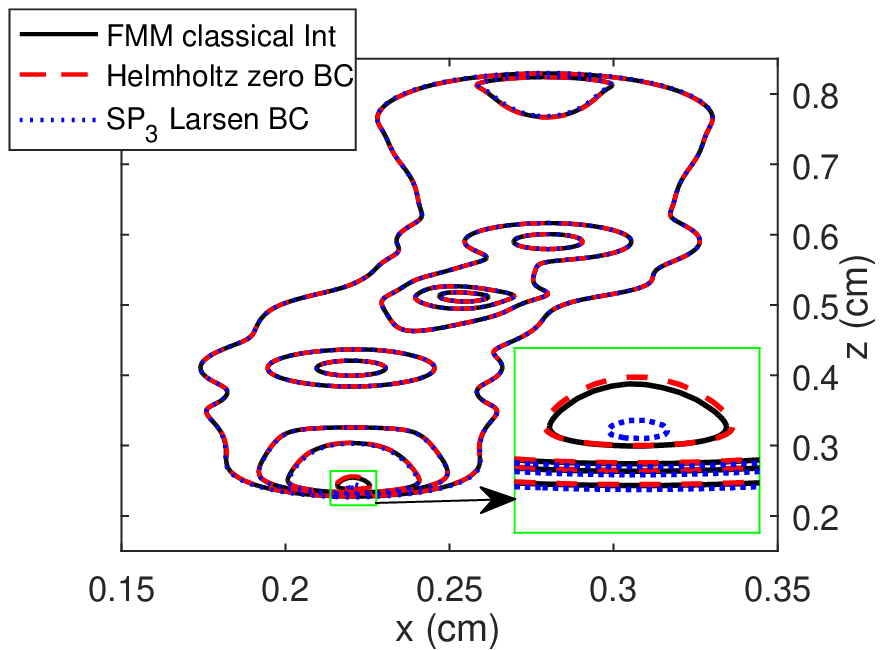}
	\includegraphics[width=0.48\textwidth]{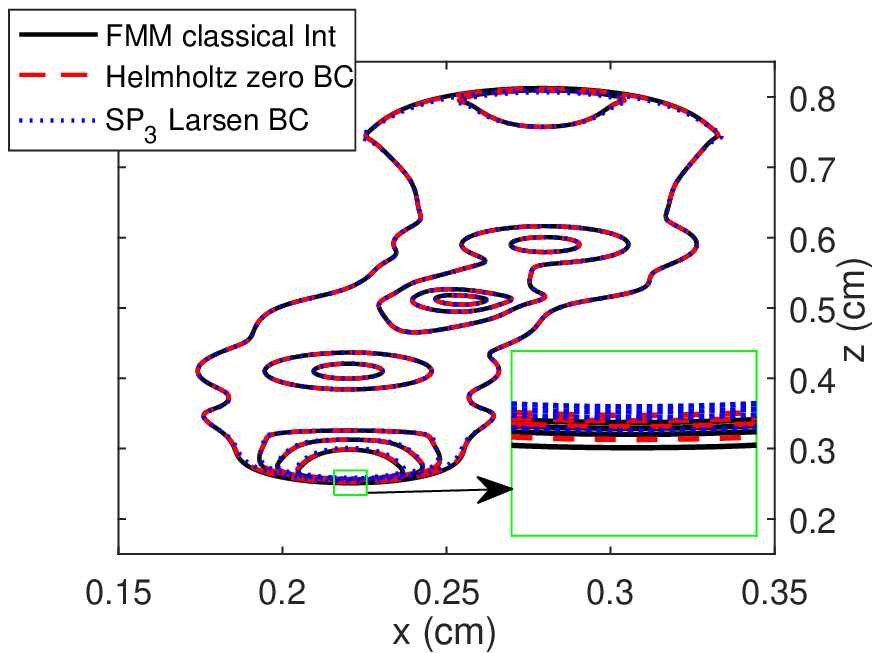}
	\caption{\label{fp760sigma}{\color{black} Contours of different electron density values at $n_e=1 \times 10^{13}$, $5 \times 10^{13}$, $9 \times 10^{13}$, $1.3 \times 10^{14}$\,cm$^{-3}$, on plane $y=0.25$\,cm at $1.5$\,ns for $p_{_{O_2}} = 150$\,Torr (left) and $p_{_{O_2}} = 0.1$\,Torr (right).} }
\end{figure}

We first compare the three methods by observing the electron density. The contours of the electron densities at 1.5\,ns are shown in Figure \ref{fp760sigma}, where the curves of different methods are plotted as different line styles and colours. Generally, the results of the three different methods are in good agreement in most part of the domain for both partial pressures of oxygen, while some differences can be observed at the heads of streamers. The differences are particularly obvious at the head of positive streamer, which is zoomed in the same figure. The generally good agreement can be attributed to a stronger influence of the impact ionization comparing to the photoionization in the region with higher electric field, while the pronounced difference at the head of positive streamer may be due to the fact that photoionization plays a more important role in the propagation of positive streamers compared with the negative ones.

Additionally, Figure \ref{fp760sigma} also displays the difference among three methods with respect to different $p_{_{O_2}}$. As expected from Section \ref{DiffP}, the difference between three methods are smaller in higher partial pressure of oxygen (150\,Torr), and more observable when $p_{_{O_2}}$ is lower (\addi{0.1\,Torr}). This implies the validity of using the Helmholtz zero BC method and SP$_3$ Larsen BC for the photoionization in higher $p_{_{O_2}}$ and also the necessity of using the FMM classical Int method in lower $p_{_{O_2}}$\addi{for long-time simulations}.

\begin{figure}[htb!]
\centering
\includegraphics[width=0.48\textwidth]{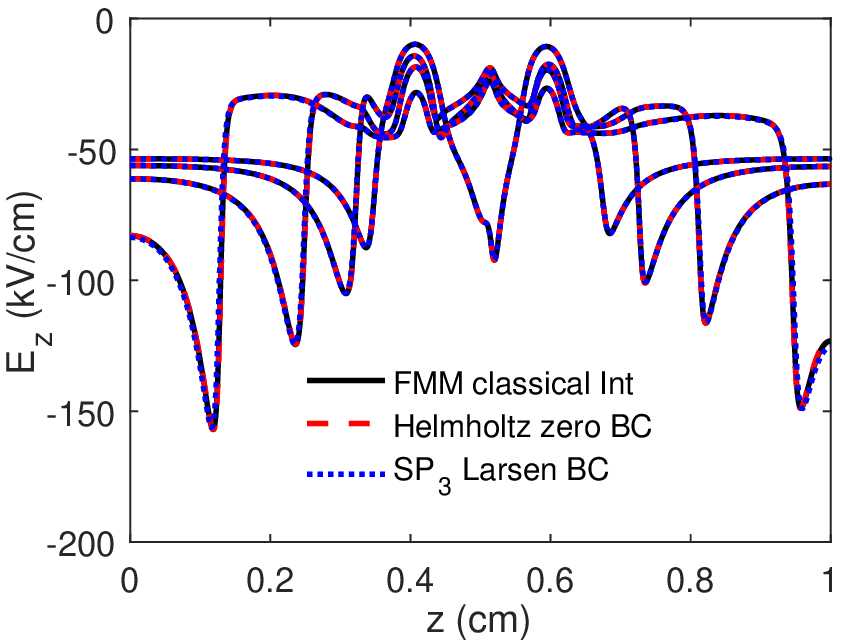}
\includegraphics[width=0.48\textwidth]{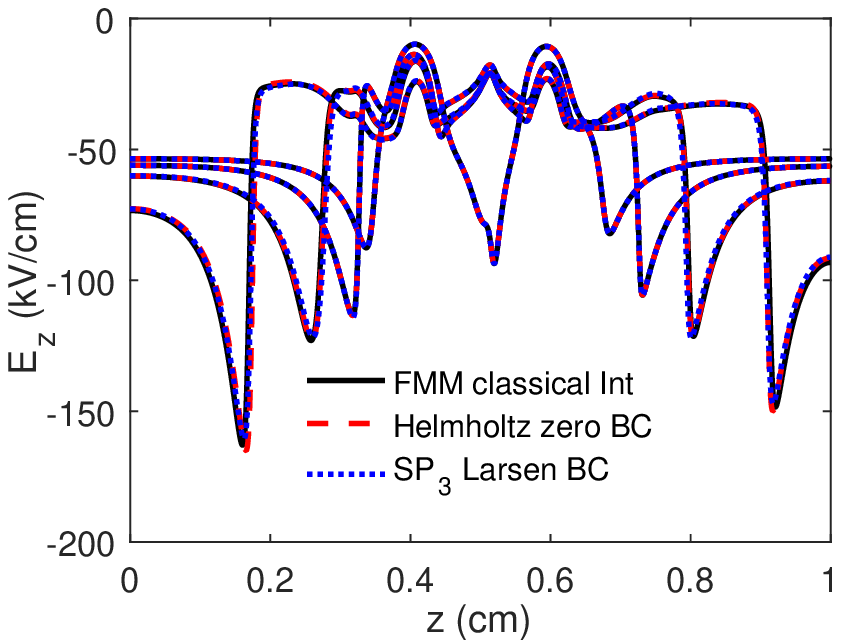}
\caption{\label{fp760Ez}{ \color{black} The third component $E_z$ of the electric field $\vec{E}$ along line $x=y=0.25$\,cm, at $0.5$, $1.0$, $1.5$ and 2.0 \,ns for $p_{_{O_2}} = 150$\,Torr (left) and $p_{_{O_2}} = 0.1$\,Torr (right). }}
\end{figure}

Besides observing the electron density at a fixed time 1.5\,ns in Figure \ref{fp760sigma}, the third component $E_z$ of the electric field $\vec{E}=(E_x,E_y,E_z)^T$ along the line $x=y=0.25$\,cm at 0.5\,ns, 1.0\,ns, 1.5\,ns {\color{black} and 2.0\,ns} is also shown in Figure \ref{fp760Ez}. As expected from Figure \ref{fp760sigma}, the differences of the three methods are generally small, while the difference are easier to be observed near the heads of streamers (the leftmost and rightmost minimum points). The difference is larger when $p_{_{O_2}}$ is small as 0.1\,Torr, and the deviation increases over time, which is consistent with the results in \cite{fvmrte2008}. One can see that, in the result of the FMM classical int method, the head of the streamer propagates slightly faster than the other two, which is possibly due to the underestimation of photoionization using the Helmholtz zero BC method and SP$_3$ Larsen BC method. As a summary, these results indicate that the accurate approximation of the photoionization could be significant in simulations with long-time propagation of streamers, especially when the partial pressure of oxygen is low.

We remark that though accurate simulation of photoionization is targeted, the accuracy might be not so significant in some cases when the applied field is strong enough and the collision ionization dominates the streamer development because the photoionization only provides the seed electrons. For these cases, the FMM could be accelerated by only considering a smaller domain. For example, one could evaluate $S_{\rm ph}$ only for a small portion of the domain where photoionization is important, and only considering those $I(\vec{y}_j)$ larger than a threshold. However, for PDE-based methods, they have to compute all the values in the domain due to the influence of boundary conditions.

\subsection{Scalability of the FMM classical Int method}\label{scalability}
As demonstrated previously, one advantage of the FMM method is the scalability in parallel computing with distributed memory, which means the ability to reduce the execution time as the number of processes increases. In this subsection, we study the scalability of the FMM classical Int method, which is quantified by the relative speed-up, defined by the ratio of the execution time using the smallest number of cores over the execution time of the parallel program.

In this test, the governing equation is again (\ref{eM}), and the initial value $\tilde{n}_0$ in \eqref{initst} is set as one Gaussian,
\[
\tilde{n}_0(\vec{x}) = 10^{14} \exp\left(- \left( (x-0.2)^2 + (y-0.2)^2 + (z-0.1)^2\right) / (0.03)^2 \right) \text{cm}^{-3}.
\]
All physics parameters are similar as those in previous subsection except stated otherwise. We set the computational domain as $[0,0.4] \times [0,0.4] \times [0,0.2]$\,cm$^3$, and adopt two uniform meshes with $256 \times 256 \times 160$ and $512 \times 512 \times 320$ grid cells. The simulation is run until $5 \times 10^{-2}$\,ns with a fixed time step $1 \times 10^{-3}$\,ns.
It should be noted that $S_{\rm ph}$ is evaluated twice in each time step, and therefore the FMM classical Int method is applied 100 times in one simulation.

\begin{figure}[htb!]
\centering
\includegraphics[width=0.45\textwidth]{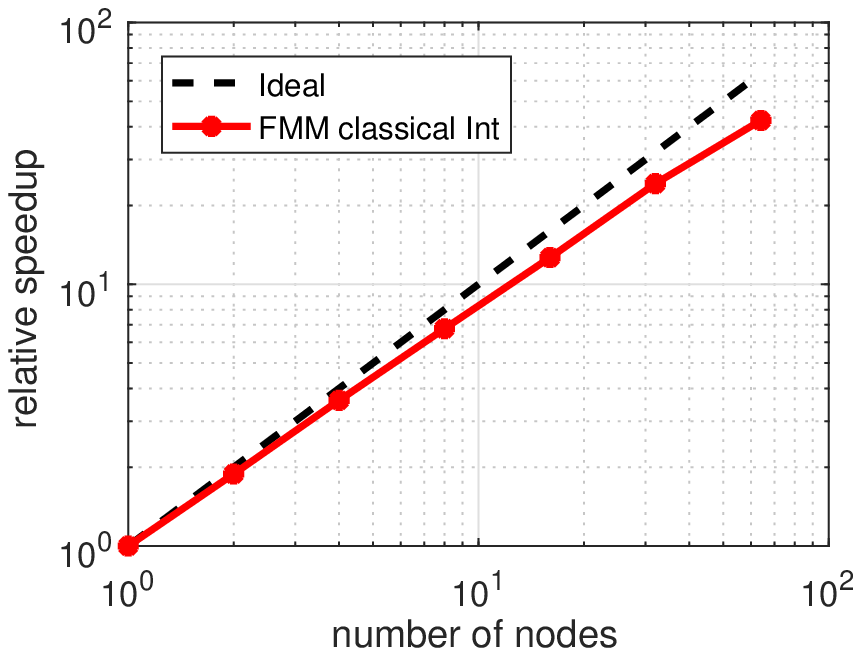}
\includegraphics[width=0.45\textwidth]{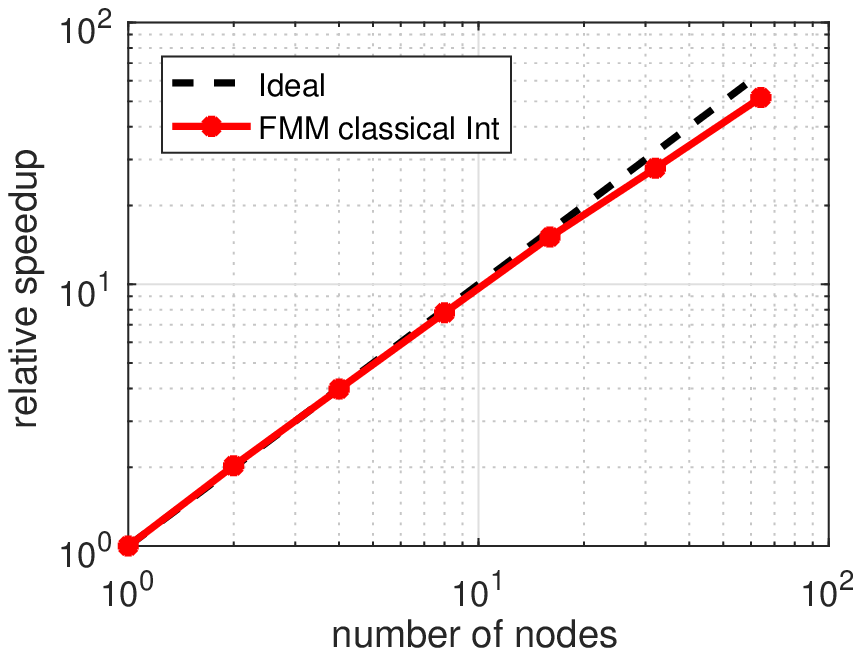}
\caption{\label{fscala}{\color{black}Relative speedups over two meshes: $256 \times 256 \times 160$ (left) and $512 \times 512 \times 320$ (right). 20 cores are used in each node.} }
\end{figure}

The time usage for the FMM classical Int method in whole simulation (100 evaluations) using different numbers of CPU cores is given in Table \ref{tscala} and plotted in Figure \ref{fscala}, where a satisfactory scalability can be observed.

\Table{\label{tscala}{\color{black}Time usage (s) using different nodes over two meshes. 20 cores are used in each node. Mesh 1: $256 \times 256 \times 160$; Mesh 2: $512 \times 512 \times 320$.}}
\br
No. of nodes & 1 & 2 & 4 & 8 & 16 & 32 & 64\\
\br
Mesh 1 & {\color{black}3828.76} & {\color{black}2031.27} & {\color{black}1062.88} & {\color{black}565.840} & {\color{black}302.655} & {\color{black}157.927} & {\color{black}90.5602} \\
Mesh 2 & {\color{black}32019.2} & {\color{black}15803.7} & {\color{black}8046.33} & {\color{black}4114.26} & {\color{black}2112.47} & {\color{black}1153.09} & {\color{black}619.401} \\
\br
\end{tabular}
\end{indented}
\end{table}

\section{Conclusion}\label{conclusion}
This paper focuses on the accurate and efficient calculation of the photoionization, and proposes the kernel-independent fast multipole method to directly compute the photoionization rate efficiently.

Quantified accuracy and time usage of the fast multipole method are studied in comparison of the classical integral model and existing approximation models based on conversion to differential equations.
The comparison shows when calculating the photoionization, the fast multipole method outperforms previous approximations in the following senses: 

(i) it is significantly efficient (or computationally cheaper) compared with the direct calculation by the classical integral under similar accuracy; 

(ii) it is much more accurate compared with those PDE-based approximations (with simple or efficient boundary conditions) under similar computational cost (same order), despite of the pressure; 

(iii) it is no need to fit additional parameters for the photoionization model, and the method is more robust with respect to domain sizes and pressures;

(iv) it is easy to be extended to unstructured meshes.

In summary, in terms of efficiency and accuracy as well as applicability to arbitrary domain with unstructured mesh, the fast multipole method
demonstrates better performance than those existing numerical methods for the calculation of photoionization in streamer discharges in the literature. 

Finally, we remark that it is straightforward to apply the kernel-independent fast multipole method to compute photoionization models with other integral forms, and thus we provide a general framework for the fast and accurate evaluation of newly proposed integral models of photoionization in streamer discharges. 

Future works include applying the method to other integral models and taking the stochastic effect into consideration.

\section*{Acknowledgments} This work was partially supported by the National Science Foundation of China under project 51921005 (R. Zeng) and 52022044 (C. Zhuang),
the Academic Research Fund of the Ministry of Education of Singapore under grants R-146-000-305-114 (Z. Cai) and R-146-000-290-114 (B. Lin and W. Bao).
Some computations were done on the Tianhe2-JK cluster at Beijing Computational Science Research Center under the kind support of Prof. Yongyong Cai.

\section*{References}
\bibliography{ssnalbib}{}
\bibliographystyle{unsrt}
%\begin{thebibliography}{10}
%\bibitem{ref1} J.~Doe, Article name, \textit{Phys. Rev. Lett.}
%
%\bibitem{ref2} J.~Doe, J. Smith, Other article name, \textit{Phys. Rev. Lett.}
%
%\bibitem{web} \href{http://www.google.pl}{www.google.pl}
%\end{thebibliography}

\end{document}